\documentclass[a4paper,11pt]{article}
\pdfoutput=1
\usepackage{mystyle}
\usepackage{graphicx}
\usepackage{color}
\usepackage{amssymb}
\usepackage{amsmath}
\usepackage{mathabx}
\usepackage{upgreek}
\usepackage{bbm}
\usepackage{multirow}
\usepackage{multicol}
\usepackage{amscd}
\usepackage{ipa}
\usepackage{mfpic}
\usepackage{verbatim}
\usepackage{hyperref}
\usepackage{cancel}
\usepackage{chemarrow}
\usepackage{tikz-cd}
\usepackage{tocloft}
\usepackage{MnSymbol} 

\definecolor{cream}{rgb}{.97, .95, .88}
\definecolor{darkcream}{rgb}{1., .88, .5}
\definecolor{lightpink}{rgb}{0.98, 0.88, 0.87}
\definecolor{lightwhite}{rgb}{1., 0.98, 0.95}
\definecolor{lightsalmon}{rgb}{1., 0.95, 0.90}
\definecolor{lightviolet}{rgb}{0.9, 0.8, 0.9}
\definecolor{lightgray}{rgb}{.96, .96, .96}  
\definecolor{lgray}{rgb}{.75, .75, .75}
\definecolor{LemonChiffon}{rgb}{0.95, 1., 0.7}
\definecolor{lightolivegreen}{rgb}{0.84, 0.89, 0.25}
\definecolor{lightgreen}{rgb}{.664, 1., .52}
\definecolor{llgreen}{rgb}{.900, .983, .960}
\definecolor{tristle}{rgb}{0.87, 0.67, 0.77} 
\definecolor{pink}{rgb}{0.95, 0.45, 0.75}
\definecolor{magenta}{rgb}{1., 0, 1.}
\definecolor{violet}{rgb}{0.9, 0.20, 0.85}
\definecolor{darkolivegreen}{rgb}{0.55, 0.65, 0.35}
\definecolor{maroon}{rgb}{0.7, 0.26, 0.56}
\definecolor{lightmaroon}{rgb}{0.85, 0.38, 0.58}
\definecolor{darkmaroon}{rgb}{0.604, 0.169, 0.451}
\definecolor{ddarkmaroon}{rgb}{0.2, 0.03125, 0.150}
\definecolor{mediumorchid}{rgb}{0.8, 0.33, 0.83}
\definecolor{mediumorchidd}{rgb}{1., 0.33, 0.63}
\definecolor{darkgreen}{rgb}{0.1, 0.6, 0.13}
\definecolor{lightyellow}{rgb}{1., 1., 0.82}
\definecolor{turquoise}{rgb}{0.042, 0.586, 0.512}
\definecolor{turquoisel}{rgb}{0.66, 0.94, 0.83}
\definecolor{darkturquoise}{rgb}{0.21, 0.55, 0.50}
\definecolor{coral}{rgb}{1., 0.6, 0.21}
\definecolor{lightorange}{rgb}{1., 0.88, 0.75}
\definecolor{orangered}{rgb}{1., 0.5, 0.}
\definecolor{orange}{rgb}{1., 0.65, 0.1}
\definecolor{orangel}{rgb}{1., .85, .3}
\definecolor{darkorange}{rgb}{0.875, 0.4, 0.204}
\definecolor{ddarkorange}{rgb}{.675, .218, .05}
\definecolor{bluesky}{rgb}{0.48, 0.53, 1.}
\definecolor{gold}{rgb}{1., 0.85, 0.25}
\definecolor{goldd}{rgb}{0.95, 0.75, 0.05}
\definecolor{darkviolet}{rgb}{0.54, 0.04, 0.84}
\definecolor{ddarkviolet}{rgb}{.382, .063, .657}
\definecolor{lightblue}{rgb}{0.30, 0.86, 0.89}
\definecolor{LightBlue}{rgb}{0.68, 0.85, 0.9}
\definecolor{lblue}{rgb}{0.78, 0.90, 0.95}
\definecolor{darkblue}{rgb}{.105, .308, .707}
\definecolor{lightmaroon}{rgb}{0.85, 0.38, 0.58}
\definecolor{darkmaroon}{rgb}{0.604, 0.169, 0.451}
\definecolor{darkpink}{rgb}{0.879, 0.020, 0.766}
\definecolor{ddarkpink}{rgb}{0.738, 0.195, 0.406}
\definecolor{grey}{rgb}{0.717, 0.717, 0.717}
\definecolor{lightgrey}{rgb}{0.800, 0.800, 0.800}
\definecolor{brown}{rgb}{0.740, 0.323, 0.182}
\definecolor{redbrown}{rgb}{.575, .158, .05}
\definecolor{darkbrown}{rgb}{0.34, 0.25, 0.05}
\definecolor{orangebrown}{rgb}{0.433, 0.262, 0.06}
\definecolor{pinkl}{rgb}{1., 0.788, 0.918}
\definecolor{salmon}{rgb}{1., 0.66, 0.5}
\definecolor{lightbrown}{rgb}{0.703, 0.508, 0.121}

\def\Journal#1#2#3#4{{#1} {\bf #2} (#3) #4}

\def\etal{{\it et al.}}
\def\Name#1#2 {{ #1 }{#2}}

\def\AA{\em A.\& A.}

\def\ACQ{\em Academia Quantum}

\def\APH{\em Annals Phys.}

\def\APJ{\em ApJ.}

\def\ASB{\em Ann. Soc. Sci. Brux. \emph{A}}

\def\CMP{\em Commun. Math. Phys.}

\def\CPH{\em Commun. Phys.}

\def\CQG{\em Class. Quant.~Grav.}

\def\EJA{{\em Europ. J. Phys.} \emph{A}}

\def\FOP{\em Found. Phys.}

\def\GRG{\em Gen. Rel. Grav}

\def\IMA{{\em Int. J. Mod. Phys.} \emph{A}}
\def\IMD{{\em Int. J. Mod. Phys.} \emph{D}}

\def\JHE{\em J. High Energy Phys.}

\def\JMP{\em J. Math. Phys.}
\def\JPA{\em J. Phys. A: Math. Theor.}

\def\JPM{\em J. Phys. Condens. Matter}
\def\JPU{\em J. Phys. (USSR)}

\def\MDU{\em MDPI Universe J.}
\def\MPA{{\em Mod. Phys.} \emph{A}}
\def\MPL{{\em Mod. Phys. Lett.} \emph{A}}

\def\NAC{\em Nature Commun.}

\def\NAT{\em Nature}

\def\NCE{\em Nuovo Cimento}

\def\NPB{{\em Nucl.~Phys.}~\emph{B}}

\def\PDU{\em Phys. Dark Univ.}

\def\PLB{{\em Phys. Lett.}~\emph{B}}

\def\PRB{{\em Phys.~Rev.}~\emph{B}}

\def\PRD{{\em Phys.~Rev.}~\emph{D}}

\def\PRX{{\em Phys.~Rev.}~\emph{X}}

\def\PRL{\em Phys. Rev. Lett.}

\def\PRV{\em Phys.~Rev.}
\def\PRX{{\em Phys.~Rev.}~\emph{X}}

\def\PTO{\em Phys. Today}

\def\RMP{\em Rev. Mod. Phys.}
\def\RPP{\em Rept. Prog. Phys.}

\def\SYM{\em Symmetry}

\def\be{\begin{equation}}
\def\ee{\end{equation}}
\def\bea{\begin{eqnarray}}
\def\eea{\end{eqnarray}}
\def\bes{\begin{equation*}}
\def\ees{\end{equation*}}
\def\beas{\begin{eqnarray*}}
\def\eeas{\end{eqnarray*}}

\def\tr{\text{tr}}

\DeclareMathVersion{normal2}
\mathversion{normal2}

\def\am{\mathcal A}
\def\bm{\mathcal B}

\def\dm{\mathcal D}

\def\fm{\mathcal F}
\def\gm{\mathcal G}
\def\hm{\mathcal H}
\def\lm{\mathcal L}

\def\nm{\mathcal N}

\def\um{\mathcal U}

\def\hA{\hat{A}}
\def\hB{\hat{B}}
\def\hF{\hat{F}}
\def\hG{\hat{G}}
\def\hH{\hat{H}}

\def\hL{\hat{L}}
\def\hO{\hat{O}}

\def\hT{\hat{T}}
\def\hU{\hat{U}}

\def\hrho{\hat{\rho}}
\def\hvarrho{\hat{\varrho}}

\def\sD{\cancel{D}}

\def\sqgr{$SU(\infty)$-QGR}
\def\suinf{{SU(\infty)}}
\def\suinfa{{\mathcal{SU}(\infty)}}

\title{Energy-momentum and dark energy in $\boldsymbol{\suinf}$-QGR quantum gravity}
\author[a,b]{Houri Ziaeepour}
\affiliation[a]{Institut UTINAM, CNRS UMR 6213, Observatoire de Besan\c{c}on, Universit\'e de Franche Compt\'e, 41 bis ave. de l'Observatoire, BP 1615, 25010 Besan\c{c}on, France}
\affiliation[b]{Mullard Space Science Laboratory, University College London, Holmbury St. Mary, GU5 6NT, Dorking, UK}
\emailAdd{houriziaeepour@gmail.com}

\abstract
{$\suinf$ Quantum GRavity (QGR) is a recently proposed fundamentally 
quantum approach to gravity and cosmology. In this model the Hilbert space of the Universe 
represents $\suinf$ symmetry. Its fragmentation generates approximately isolated subsystems 
(particles) representing, in addition to $\suinf$, finite-rank local symmetries. The common 
$\suinf$ is associated to quantum gravity, and at lowest quantum order the effective action for 
all symmetries is Yang-Mills on a 4D parameter space $\Xi$. Nonetheless, physical processes and 
measurables must be independent of the geometry of $\Xi$. In previous works we demonstrated that 
diffeomorphism of $\Xi$ can be neutralized by $\suinf$ gauge transformation. In this work we show 
that the invariance of action under variation of $\Xi$'s metric leads to a constraint resembling 
Einstein equation. It consists of energy-momentum tensors for all components of the model, 
including the spin-1 gravitons. In addition, through calculation of quantum information measures 
we study the effect of Hilbert Space Fragmentation (HSF) on the emergent classical spacetime 
and different cosmological era, such as inflation, reheating, and late time 
accelerating expansion. The results of this preliminary and approximate investigation show that 
HSF may be classically interpreted as these phenomena. Consequently, inflaton, quintessence, and 
similar fields associated to these processes may be order parameters that phenomenologically 
present them.
}

\begin{document}

\maketitle
\flushbottom


\section{Introduction} \label{sec:intro}
We live in a Universe ruled by laws of quantum mechanics. All entities 
(particles), forces (interactions) and their observables are formulated in this framework. 
Although, once upon the time it was assumed that quantum mechanics only applies to microscopic 
scales, progress in theoretical and experimental physics has shown that quantum physics is needed 
for explaining all aspects of our Universe. Examples of macroscopic objects behaving quantum 
mechanically include: Quantum coherent nanoparticles producing quantum 
interference~\cite{qmnanoparticle}; cores of neutron stars, which are analogous to nucleus of a 
giant atom~\cite{nstarqcdrev} (review); large scale anisotropies in the Universe, best explained 
by quantum fluctuations of fields in the early Universe~\cite{qftinfplanck}. Even properties which 
apparently seem classical can and need to be explained by quantum 
mechanics~\cite{qmdecohereclassic,qmdecohereclassic0}, because the underlying processes have 
quantum origin.

The exception is what we perceive as spacetime and its geometry felt as gravitational force.
Although the spacetime and gravity apparently behave in a deterministic manner, they are closely 
related to quantum matter, which is the source of gravity. The inconsistency of such situation 
is well known, see for example~\cite{qmgrinconsist,qmgrinconsist0,qmgrinconsist1,qmgrinconsist2}. 
However, so far quantization gravity through many methods and models have not led to a fully 
consistent and testable quantum gravity model. Examples of some popular attempts 
include\footnote{Citation in this paragraph are indicative and are very far from being exclusive.}: 
canonical quantization of the Einstein-Hilbert action as a field theory~\cite{admgr,qgrcanonical}; 
quantization of curved spaces using triangulation~\cite{curvatureregge,qgrponzanoreggge} and 
related ideas such as loop QGR~\cite{lqgrev,lqgrev0} (reviews), spacetime 
foam~\cite{lqgfoam,qgrspinfoam,qgrspinfoam0}, group field theory~\cite{lqggroup} (review); 
introduction of extended entities such as strings and branes~\cite{stringrev,stringbranerev} 
(reviews) as fundamental quantum entities; quantum entanglement as 
gravity~\cite{qgrentangle,qgrentangle1}. In addition, there are stringent constraints on the 
rare predictions of these models, such as modification of dispersion relation of relativistic 
particles by micro-structure of spacetime~\cite{grestgrb090510a,qgrdispergrb}, violation of 
Lorentz invariance~\cite{lorentzviolconstr,lorentzviolconstr0,lorentzviolconstr1}, modification 
of general relativity~\cite{grtestligo}, existence of large 
extra-dimensions~\cite{braneuhecr,branecmblss,branecmblss0,branecmblss1,branecmblss2}, 
and existence of a large number of massless or light particles called 
modulies~\cite{cmbplancknunum}. Moreover, the cited models do not explain the origin and physical 
nature of spacetime and its dimensionality. Motivated by the progress in Quantum Information 
Theory(QIT) and by gauge-gravity duality conjecture~\cite{gaugegr,qgrgaugesep}, especially 
Anti-de Sitter - Conformal Field Theory (AdS-CFT) conjecture~\cite{adscftrev} (review), which is 
proved in (2+1)D spacetimes\cite{adscftproof}, a few inherently quantum models are proposed. 
The \sqgr~is one of them.

\sqgr~is based on an abstract approach to the Universe, its contents, and their interactions. 
It diverges from other QGR candidates by the fact that it is fundamentally quantum and is not 
related to any classical model. Moreover, there is no spacetime or gravity in its construction. 
These entities emerge from abstract axioms of the model and properties of the Universe as 
an isolated quantum system. In particular,  the classical (3+1)-dimensional spacetime emerges 
as effective (average) parameters that characterize quantum states of the contents of the 
Universe's. In a QIT view, the latter corresponds to approximately isolated quantum subsystems 
that emerge as a result of the fragmentation of the Hilbert space of the Universe. The model 
was first proposed in~\cite{houriqmsymmgr} and its  properties are investigated 
in~\cite{,hourisqgr,hourisqgrym}. It is reviewed in~\cite{hourisqgrhighlight,hourisqgressay}. 

In this work we first address the issue of energy-momentum tensor and its physical 
interpretation in the framework of \sqgr. We also investigate what may be called {\it vacuum 
energy density} in this framework, and whether it can play the role of a Cosmological Constant 
(CC), as it was introduced by Einstein~\cite{einsteincc} and interpreted as {\it the energy 
density of the vacuum} by Lema\^itre~\cite{lemaitrevacener}. We address dynamics of the quantum 
state of the Universe and whether and how it may behave as a dynamical dark energy, as the 
recent cosmological observations may indicate~\cite{dedynamical}.

As mentioned earlier, in \sqgr~the classical spacetime $\Sigma$ is not a physical entity, but a 
property that presents the effective or average path of the subsystems - contents - of the 
Universe in their (3+1)D space $\Xi$ of continuous parameters, which characterize their quantum 
states. As expected from the space of parameters, geometry of $\Xi$ should not be an observable 
and one should be able to redefine parameters. On the other hand, in general relativity 
energy-momentum tensor $T_{(m)}^{\mu\nu}$ of matter is defined based on the functional 
derivative of its action with respect to the metric of spacetime and its derivative. Therefore, 
following questions arise: Which metric should we use for the definition of observable 
$T_{(m)}^{\mu\nu}$ ? Should we consider this tensor as an effective or average classical quantity 
or a corresponding quantum operator can be associated to it ? And in the latter case how should 
we tackle the well known difficulties in the renormalization of energy-momentum tensor and the 
related issue of vacuum energy. In this work we address these issues.

Sec. \ref{sec:vacenermom} provides a physical interpretation for the effective Lagrangian 
functional of \sqgr~obtained in the previous works~\cite{houriqmsymmgr,hourisqgr}, and based 
on this interpretation we define a Lagrangian superoperator for the model. Moreover, we show 
that the independence of the Lagrangian from metric of the parameter space results to a 
constraint equation. For each field species in the Lagrangian the corresponding term in the 
constraint equation has a similar expression as the energy-momentum tensor in general 
relativity, but up to a sign for $\suinf$ Yang-Mills field, namely the quantum gravity. These  
results provide a concrete definition for gravitational energy-momentum tensor, which does not 
exist in the classical case~\cite{grwaldbook}. We remind that in general relativity the concept 
of energy-momentum tensor of gravity is addressed only for fluctuations of the metric around a 
background in the context of gravitational waves. Specifically, in order to employ the usual 
definition of energy-momentum tensor, fluctuations of metric should be treated as a spin-2 
{\it matter}. We use these results to define a Hamiltonian superoperator that can be used to 
study the evolution of quantum states of subsystems. Sec. \ref{sec:desqgr} discusses the 
issue of dark energy. In~\cite{hourisqgr} we proposed a few \sqgr~specific models for dark 
energy. Here, along a discussion about the concept of vacuum energy in \sqgr~we briefly and in 
a critical manner review these models and assess their plausibility. Moreover, we use the 
relationship between an affine separation, associated to average path of the quantum state of 
the Universe, to investigate the effect of state evolution on the effective metric perceived as 
that of the classical spacetime. We explore how the variation of quantum state can affect 
phenomena such as inflation and accelerating expansion of effective metric. Finally, 
Sec. \ref{sec:outlines} summarizes the results and conclusions of this work.

Supplementary sections contain reviews and details of calculations discussed in the main text. 
Sec.\ref{app:sqgrrev} briefly reviews \sqgr, in particular those properties of the model 
important for the present work. Thus, we assume that readers have sufficient familiarity with 
the model. The proposal of fragmentation of the quantum state and Hilbert space of the Universe 
is motivated by analogous phenomena in isolated many-body systems in condensed matter. For this 
reason we review these processes in Sec. \ref{app:mbnonthermal}. Two QIT measures are useful for 
quantifying the emergence of clustering in quantum states: coherence and state overlap. In 
Sec. \ref{app:decohere} we calculate change of coherence for completely coherent states under 
application of $\suinf$ generators. In Sec. \ref{app:stateoverlap} we determine evolution of 
overlap between an arbitrary state and its variation under successive application of $\suinf$ 
generators.

\section{Lagrangian and Energy-Momentum Tensor in $\boldsymbol{SU(\infty)}$-QGR}  \label{sec:vacenermom}
As explained in Sec. \ref{app:sqgrrev} in \sqgr~there is no {\it background spacetime} and what 
is perceived as the classical spacetime presents average or effective values of the continuous 
parameters characterizing the path of quantum states of subsystems in their Hilbert space. 
Consequently, the Lagrangian functionals for the whole Universe and its subsystems defined in 
Sec. \ref{app:sqgrrev} and energy-momentum tensor have interpretations somehow different 
from their analogues in general relativity and QFT. In this section we discuss these important 
topics, before considering the much more complicated issue of the nature of observed 
dark energy in the next section.

\subsection{Physical Interpretation of Lagrangian Functionals}  \label{sec:lagranginterprt}
We should first remind that \sqgr~is constructed as a quantum model. Therefore, although 
Lagrangians $\lm_U$ and $\lm_{U_s}$ defined in (\ref{lagrange2d}) and (\ref{yminvarsub}), 
respectively are C-numbers, unlike classical Lagrangians used in QFT on an external spacetime 
they have quantum origin. 

We begin with the Lagrangian functional $\lm_U$ in (\ref{lagrange2d}) constructed on the 
Hilbert space of the whole Universe. Considering the Lagrangian density $\lm_U (\theta,\phi)$, 
the tracing operation is equal to summing square of eigen values of $\suinf$ generators 
$\hL_a (\theta,\phi)$. Indeed, parameters $(\theta,\phi)$ characterize generators of $\suinf$. 
Thus, integration over coordinates of the diffeo-surface $D_2$ is the summation over all 
generators. The $\suinf$ field strength $F_a^{\mu\nu} (\theta,\phi),~\mu,~\nu = 0,~1$ is 
amplitude of the corresponding generator in the state - the wave function - of the Universe. 
Hence, $F^{\mu\nu}F_{\mu\nu}$ is the probability of pointer state characterized by $(\theta,\phi)$. 
Although the second term in $\lm_U (\theta,\phi)$ looks different, in what concerns $\suinf$ 
representation it is similar to the first term, because the state 
$\hvarrho_U \in \bm[\hm_U] \cong \suinf$, it can be expanded with respect to 
$\hL_a (\theta,\phi)$ generators. The reason for separating this term is that in analogy with 
the matter term in Yang-Mills QFT, 
$\hvarrho_U (\theta,\phi) \equiv |\Psi_U\rangle \langle \Psi_U|$ and 
$|\Psi_U (\theta,\phi)\rangle$ can be in a different representation of the Lorentz symmetry 
of the parameter space $D_2$. Specifically, $F^{\mu\nu}$ in the first term is a 2-form on $D_2$, 
whereas $|\Psi_U (\theta,\phi)\rangle$ can be in another representation, for instance a spinor, 
especially when the state is fragmented to subsystems. 

Interpretation of Lagrangian $\lm_{U_s}$ in (\ref{yminvarsub}) is similar to that of the whole 
Universe. It presents a 1-subsystem (particle) effective Lagrangian at the lowest quantum order. 
An important feature of $\suinf$ Yang-Mills model presented in the Lagrangian $\lm_{U_s}$ is 
the dependence of what is called {\it internal coordinates}~\cite{suninfym}, that is coordinates 
of the diffeo-surface $(\eta, \zeta)$ of the $\suinf$ representation realized by the subsystem. 
In contrast to $\suinf$ Yang-Mills on a background spacetime they are not independent from 
coordinates of the parameter space $\Xi$~\cite{hourisqgrym}. Therefore, interpretation of 
amplitudes $F_a^{\mu\nu} (\theta,\phi),~\mu,~\nu = 0, \cdots,~3$, trace operation, and integration 
over these parameters in $\lm_{U_s}$ are the same as those given in the previous paragraph for 
the Lagrangian functional of whole Universe.

\subsubsection {Lagrangian Superoperator}  \label{sec:lagrangeop}
The Lagrangian functionals $\lm_U$ and $\lm_{U_s}$ are real numbers. Nonetheless, using 
the above interpretation of these functionals in the framework of \sqgr~we can also define 
{\it Lagrangian super operators} which act on $\bm[\hm_U]$ of the whole Universe, that is 
$\hat{\lm}_U: \bm[\hm_U] \rightarrow \bm[\hm_U]$ and that of the subsystems: 
$\hH_i: \bm[\hm_i] \rightarrow  \bm[\hm_i]$. Here the index $i$ indicate a specific subsystem. 
Using (\ref{lagrange2d}) and \ref{yminvarsub} we define these operators such that their 
expectation values be equal to the C-number Lagrangians $\lm_U$ and $\lm_{U_s}$, respectively. 

For the whole Universe at lowest order the Lagrangian superoperator is defined as:
\be
\hat{\lm}_U \equiv \int d^2\Omega ~\hat{\lm}_U (\theta,\phi) \equiv \int d^2\Omega ~ 
\biggl [ ~\frac{1}{2} ~ \hF^{\mu\nu} \hF_{\mu\nu} + \frac {1}{2} \sD \hvarrho_U \biggr ], \quad 
\mu,~\nu = 0,~1 \label{oplagrange2d} 
\ee

Here we have added a symbol ``$~\hat{}~$'' over fields to remind that they are members of 
$\bm[\hm_U]$.

For subsystems representing a finite rank {\it internal} symmetry $G$ the Lagrangian 
superoperator at lowest order can be written as:
\be
\hat{\lm}_{U_s} = \int d^4x \sqrt{|\upeta|} ~ \biggl [\frac{\kappa}{4} (\hF^{\mu\nu} \hF_{\mu\nu}) + 
\frac{1}{4} \hG^{\mu\nu} \hG_{\mu\nu} + \frac {1}{2} \sum_s \sD \hvarrho_s) \biggr ], 
\quad  \mu,~\nu = 0, \cdots,~3 \label{opyminvarsub}
\ee

The main difference between (\ref{oplagrange2d}) and (\ref{opyminvarsub}) and C-number 
functionals (\ref{lagrange2d}) and \ref{yminvarsub} is the absence of trace operator the formers.
Note that these operators are 1-point, that is they contain multiplication of the two $\hL$'s 
at the same point of the parameter space. For this reason there is only one integration over 
the volume of the corresponding parameter space. The 2-point terms, that is when $\hL$'s are 
applied at different points of the parameter space are equivalent to adding propagators to the 
description of effective Lagrangian as in 2-Particle Irreducible (2PI) perturbation 
theories~\cite{2pirev} (review). Such terms are of higher quantum perturbation order and can 
be ignored in the lowest order Lagrangians considered here.

\subsection{Energy-Momentum Tensor in \sqgr}  \label{sec:enermom}
As explained in Sec. \ref{app:evolsubsys}, the Lagrangian (\ref{yminvarsub}) of subsystems of 
the Universe is independent of the curvature of the (3+1)D parameter space $\Xi$ on which it 
is defined. Therefore, the metric $\upeta_{\mu\nu}$ of this space is arbitrary. This observation 
raises a question about the fate of the energy-momentum tensor $T_{\mu\nu}$, which is 
important for physics and is a measurable quantity, in the framework of \sqgr. 

In classical general relativity and QFT in curved spacetime the energy-momentum tensor of 
matter is defined as:
\vspace{-9pt}
\be
T^{\mu\nu}_{(m)}(x) \equiv \frac{2}{\sqrt{|\upeta|}} \frac{\biggl (\partial 
\sqrt{|\upeta|}\lm_{(m)}(x) \biggr )}{\partial \upeta_{\mu\nu}(x)}  \label{enermommatter} 
\ee
where $\lm_{(m)}(x)$ is the effective matter Lagrangian density and $x$ indicates a point of 
the space on which the Lagrangian is defined~\cite{renormadiab3}. In QFT in curved spacetime 
this space is an Einstein-Riemannian manifold~\cite{curvaturfunc,grwaldbook}. For calculating 
the total energy-momentum tensor, which constitutes the matter side the Einstein equation the 
Lagrangian density $\lm^{\mu\nu}_{(m)} (x)$ must include all contents of the model that are 
considered to be sources of gravitation, including the cosmological constant. Moreover, 
the energy-momentum tensor must be locally conserved, that is $D_\mu T^{\mu\nu}_{(m)} = 0$, where 
$D_\mu$ is the covariant derivative with respect to $x$. On the other hand, the Einstein-Hilbert 
action is designed such that the resulting field equation satisfy this constraint and its field 
equation be the Einstein equation~\cite{grwaldbook}.

The Einstein equation is nonlinear in metric field. This means that gravity field source itself. 
However, in general relativity there is no convincing definition for gravitational energy or 
energy-momentum tensor, see e.g. chapter 11 of~\cite{grwaldbook}. For gravitational waves 
and proof of their propagation a gravitational energy-momentum tensor is defined 
perturbatively. Specifically, the metric is decomposed to a background component - usually an 
asymptotically flat, or in the case of inflation, a de Sitter metric - and a perturbation 
around the background. This perturbative component is treated like a spin-2 matter field 
and its energy-momentum tensor is calculated according to (\ref{enermommatter}), see 
e.g.~\cite{gwrev} for a review. Indeed, in contrast to dynamic equations for non-gravitational 
fields, which are obtained from functional derivation of their action with respect to the field 
and its derivative, Einstein equation is obtained from variation of the Einstein-Hilbert action 
only with respect to the metric~\cite{grwaldbook}. This is because $D_\sigma\upeta_{\mu\nu} = 0$. 
For this reason, if we extend the definition of energy-momentum tensor (\ref{enermommatter}) 
to gravity, the geometric side of the Einstein equation can be interpreted - up to a sign - 
as {\it gravitational energy-momentum tensor}, which neutralizes that of matter through Einstein 
equation. 

In \sqgr~curvature of the parameter space $\Xi$ can be gauged out~\cite{hourisqgr,hourisqgrym}. 
However, the Lagrangian (\ref{yminvarsub}) remains depends on the metric. This can be seen as 
the remnant of the gauge freedom which must be fixed by the following additional condition:
\be
\frac{\partial \biggl (\sqrt{|\upeta|}\lm_{U_s}(x) \biggr )}{\partial \upeta_{\mu\nu}(x)} = 0, 
\quad \forall x \in \Xi  \label{lagrangiangcond}
\ee

This relation imposes 6 constraints - the number of independent components of $\upeta_{\mu\nu}$ - 
on the infinite number of $F^{\mu\nu}_a$ amplitudes of $\suinf$ (gravity) gauge field. Therefore, 
although (\ref{lagrangiangcond}) looks like the definition of Einstein equation,, it cannot be 
considered as field equation, but rather a consistency condition. On the other hand, up to an 
irrelevant $2 / \sqrt{|\upeta|}$ factor, the l.h.s. of (\ref{lagrangiangcond}) has the same 
expression as the definition of energy-momentum tensor (\ref{enermommatter}). Specifically, 
using the effective action of subsystems (\ref{yminvarsub}) and assuming a fermionic state for 
the subsystems, the constraint (\ref{lagrangiangcond}) (including the factor 
$\frac{2}{\sqrt{|\upeta|}}$) leads to:
\bea
&& \kappa \biggl (\upeta_{\rho\sigma} \tr (F^{\mu\rho} F^{\nu\sigma}) - \frac{1}{4} 
\upeta^{\mu\nu} \tr (F_{\rho\sigma} F^{\rho\sigma}) \biggr ) + 
\biggl ( \upeta_{\rho\sigma} \tr (G^{\mu\rho} G^{\nu\sigma}) - 
\frac{1}{4} \upeta^{\mu\nu} \tr (G^{\rho\sigma} G_{\rho\sigma}) \biggr ) + \nonumber \\
&& \quad \quad \quad 
\frac {1}{2} \sum_s \tr (\gamma^0 \gamma^i e_i^\mu \upeta^{\nu\sigma}
\overleftrightarrow {D}_\sigma \hrho_s) = 0  \label{enermomcond}
\eea
where the trace operators are applied to generators of $\suinf$ and $G$ symmetries. It is 
straightforward to verify that terms in the first bracket of the l.h.s. of (\ref{enermomcond}) 
correspond to$\kappa$ times of $\suinf$ - gravity - Yang-Mills field energy-momentum tensor 
$T^{\mu\nu}_{(\suinf)}$; the second bracket is equal to the energy-momentum tensor of the internal 
symmetry $G$ Yang-Mills field; and the last term is the energy-momentum tensor for {\it matter} 
state. Thus, (\ref{enermomcond}) implies:
\be
\kappa T^{\mu\nu}_{(\suinf)} + T^{\mu\nu}_{(G)} + T^{\mu\nu}_{(m)} = 0. \label{enermomequilib}
\ee

We can also use the superoperator Lagrangian (\ref{opyminvarsub}) to obtain an analogous 
constraint in operator form acting on the Fock space of the fields. An important conclusion 
from this relation is that the zero-modes of the fields, which are divergent and not 
renormalizable~\cite{zeromodecc} should cancel each others. Therefore, it is fully legitimate 
to use normal ordering and get rid of them. 

\subsubsection{Physical Interpretation}  \label{sec:enermominterprt}
If the constant factor $\kappa$ in the Lagrangian (\ref{yminvarsub}) is negative, the metric 
independence condition (\ref{lagrangiangcond}) means that in \sqgr~there is an equilibrium 
between  gravitational energy-momentum tensor and those of subsystems. In other words, 
the minimum of the effective Lagrangian $\lm_{U_s}$ corresponds to the equality between 
energy-momentum tensor of gravitational and matter fields - including finite rank gauge fields. 
This conclusion is analogous to the equilibrium between kinetic and potential energy in 
classical physics and consistent with canonical definition of Lagrangians. The negative sign 
of $\kappa$ indicates a $\pi/2$ phase difference between amplitude or generators of 
$\suinf$ (gravity) and other fields (subsystems).This global phase can be associated to 
the $U(1)$ symmetry discussed in Sec. \ref{app:subsys}. 

In place of applying the metric independence constraint to the effective Lagrangian, we could 
use the Lagrangian superoperator (\ref{opyminvarsub}). Giving the fact that $\upeta_{\mu\nu}$ 
is C-number, it is straightforward to check that the expectation value of the resultant equation 
would be the same as \ref{enermomcond}) and (\ref{enermomequilib}). The advantage of 
using the effective Lagrangian is that in the classical limit as defined in 
Sec.\ref{app:classiclimit}, the first term in \ref{enermomcond}) becomes the gravity side of 
the Einstein equation. 

To get a better physical insight into these properties we remind that in general relativity 
the Einstein-Hilbert action is designed such that geometry side of the Einstein equation can be 
obtained from its derivation with respect to the metric~\cite{grwaldbook}. Whereas, the same 
operation on the matter Lagrangian is interpreted as its energy-momentum tensor. However, if 
we consider gravity as a field similar to other contents of the Universe, according to the 
definition of energy-momentum tensor in (\ref{enermommatter}) the geometric part of the 
Einstein equation should be interpreted as energy-momentum tensor of graviton field. In 
\sqgr~the constraint (\ref{lagrangiangcond}) is not the field equation for gravity, because 
gravity field is the $\suinf$ gauge field $A_\mu$ defined in (\ref{yminvardefsuinf}). 
The constraint (\ref{enermomcond}) shows that  the general definition of matter 
energy-momentum tensor (\ref{enermommatter}) applies also to gravity. Therefore, \sqgr~provides 
a consistent definition for the missing energy-momentum tensor of gravity in classical general 
relativity. 

The constraint (\ref{lagrangiangcond}) can be also applied to the 2D geometry of $D_2$ 
parameter space of the whole Universe defined in Sec. \ref{app:axiom} and leads to 
$T^{\mu\nu}_{(U)} = 0, ~ \mu, \nu = 1,2$. This means, as expected, the whole Universe does 
not have energy or momentum. These quantities are defined with respect to an external 
reference frame, which does not exist for the whole Universe, because everything is part 
of the Universe and there is no independent external reference frame.

In summary, in \sqgr~gravity has both field equation and energy-momentum tensor. They 
are obtained in a similar manner as those of other quantum fields. By contrast, Einstein 
equation is not a field equation, but a constraint. Its emergence from a constraint confirms 
the interpretation of the Einstein equation as an {\it equation of state} 
in~\cite{einsteineqeof}, which is obtained in the framework of classical physics, using 
completely different assumptions than those of \sqgr. This convergent may have deeper physical 
and/or mathematical origin, worth for further, especially on the way to a quantum definition 
for black holes.

\subsubsection{Hamiltonian Operator}  \label{sec:hamilton}
Dynamics and observables of perturbative QFTs are usually studied through their Green's 
functions, that is the multi-point expectation value of fields. However, complex many-body 
phenomena, such as multipartite entanglement, phase transition, etc. usually study the 
evolution of multipartite density matrix.  As \sqgr~uses various concepts from these approaches 
to complex quantum systems, it is useful to calculate quantum Hamiltonian superoperator by 
which quantum state of gravity and subsystems unitarily evolve. 

Hamiltonian superoperator corresponds to $\hT^{00}$ component of the energy-momentum 
tensor operator~\cite{grwaldbook}. As explained earlier, it can be determined by applying 
the general definition of energy-momentum tensor (\ref{enermommatter}) to the Lagrangian 
operator (\ref{opyminvarsub}). In the same way as in (\ref{enermomequilib}) we distinguish 
three components in the total Hamiltonian:
\bea
\hH_{(\suinf)} & = & |\kappa| \biggl (\upeta_{\rho\sigma} \hF^{0\rho} \hF^{0\sigma} - \frac{1}{4} 
\upeta^{00} \hF_{\rho\sigma} \hF^{\rho\sigma} \biggr ) \label{grhamiltonian} \\
\hH_{(G)} & = & \upeta_{\rho\sigma} \hG^{0\rho} \hG^{0\sigma} - 
\frac{1}{4} \upeta^{00} \hG^{\rho\sigma} \hG_{\rho\sigma} \label{symmhamiltonian} \\
\hH_{(m)} & = & \frac {1}{2} \sum_s \gamma^0 \gamma^i e_i^0 \upeta^{0 \sigma}
\overleftrightarrow {D}_\sigma \hrho_s  \label{matterhamiltonian}
\eea
However, due to the dependence of all components on the $\suinf$ (gravity) field, they are not 
independent, unless gravity can be ignored.

\section{Dark Energy in $\boldsymbol{SU(\infty)}$-QGR} \label{sec:desqgr}
The Lagrangian $\lm_{U_s}$ in (\ref{yminvarsub}) does not include a cosmological constant. 
However, cosmological observations show without doubt that radiation (relativistic particles), 
and visible and dark matter cannot explain the apparently accelerating expansion of the Universe 
according to the classical general relativity, and an entity behaving, at least approximately, 
similar to a cosmological constant - often interpreted as {\it vacuum energy}.- must be present. 
In addition, recent observations of the power spectrum of the Large Scale Structure (LSS) of the 
Universe and the Baryonic Acoustic Oscillation (BAO)~\cite{hubbletensiondesi}, 
the power spectrum of Cosmic Microwave Background (CMB) anisotropies~\cite{cmbplancknunum}, and 
supernovae standard candles~\cite{hubblesn,hubblesn0} may point to a dynamical content, 
generally called dark energy rather than a cosmological constant~\cite{hubbletensionrev}. 

In this section after describing what can be considered as {\it vacuum} in \sqgr, we review a 
few dark energy models specific to \sqgr~proposed in~\cite{hourisqgr}. We further 
discuss their relationship and plausibility, especially if future observations confirm dynamical 
nature of dark energy. Contemplating such possibility, we then use the relationship between 
variation of affine separation and quantum state of the Universe to investigate their 
consequences for the effective classical metric (\ref{effmetric}) and apparent accelerating 
expansion of the Universe.

\subsection{Vacuum and Its Energy Density}   \label{sec:vacuum}
Although recent observations might indicate a dynamical dark energy, its equation of state 
remains very similar to a cosmological constant, historically interpreted as the energy of 
the vacuum~\cite{ccvacuum}. For this reason we begin by exploring this concept and its 
meaning in \sqgr.

In \sqgr~there is no vacuum ! This is a consequence of the third axiom of the model, which  
posits that there are infinite number of observables in the Universe. Thus, the Universe is 
never empty, otherwise it becomes nonexistent. This is in contrast to QGR models in which a 
background spacetime is omnipresent, irrespective of the content of the Universe. Moreover, 
scale invariance and topological nature of the action (\ref{lagrange2d}) for the whole Universe 
means that there is no constraint on the strength of $\suinf$ quantum Yang-Mills field, 
except that it cannot be null for all values of $(\theta,\phi)$. In addition, as explained in 
the previous section, in absence of time there is no $T^{00}$ and no {\it energy of vacuum}. On 
the other hand, interpretation of constraint (\ref{lagrangiangcond}) in Sec. \ref{sec:hamilton} 
showed that for the Universe fragmented to subsystems zero-modes of various contents of the 
Universe must cancel each others.Thus, in the framework of \sqgr~the observed dark energy 
cannot be the energy of a global vacuum.

Regarding the topology of diffeo-surface $D_2$, a priori we should sum over all topologies. 
However, giving the fact that $\suinf$ is homomorphic to smooth area preserving diffeomorphism, 
topology of $D_2$ cannot be changed by $\suinf$ interaction. For this reason, through this work 
we consider it as fixed, unless explicitly indicated. It is nonetheless possible to define a Fock 
space $\fm$ of {\it quantum universes}:
\be
\fm = \text{span}_{g_n,~n \in \mathbb {Z^+}}\biggl \{\hm^{g_1}_U \times \hm^{g_2}_U \ldots 
\times \hm^{g_n}_U \biggr \}  \label{univfockspace}
\ee
where $\hm^{g}_U$'s are Hilbert spaces of universes, and $g$ indices indicate genus of  
diffeo-surface of their $\suinf$ representation. As each $\hm^{g}_U \cong \suinf$, and 
$\suinf^n \cong \suinf ~ \forall n$, we conclude that $\fm \cong \hm_U$. Therefore, in the 
framework of \sqgr~{\it multiverses} are not distinguishable. Nonetheless, one subspace of 
$\fm$ is distinguishable from others: {\it the vacuum}. It corresponds to the trivial 
representation of $\suinf$. As the Universe we live in is far from being empty or trivial, the 
vacuum state has only conceptual application. In particular, it can be used to define creation 
$a^\dagger_k$ and annihilation $a_k$ operators for quantum systems representing $\suinf$ 
non-trivially. Here the index $k$ indicates the ensemble of parameters which discriminate 
these systems. 

As we discussed in Sec. \ref{app:subsys}, after approximate fragmentation of $\hm_U$, 
approximately isolated subsystems represent $\suinf \times G$ symmetry, where $G$ is a generic 
finite rank symmetry group. It distinguishes $\suinf$ representing subsystems from each others 
even in absence of information about the genus of diffeo-surface of their $\suinf$ 
representation.

\subsection{Previously Proposed Models for Dark Energy}  \label{sec:deoldmodels}
Many dark energy models in the literature are relevant in the framework of \sqgr. Nonetheless, 
in~\cite{hourisqgr} we outlined three phenomena related to the Yang-Mills nature of gravity 
in \sqgr~and its other features to show that at classical limit - as defined in 
Sec. \ref{app:classiclimit} - these processes may behave similar to dark energy: 

\begin{enumerate}
\item {\bf A static $\suinf$ field:} In analogy with static electromagnetism the field strength 
$F^{\mu\nu}$ of quantum gravitons can include a zero mode generated by all subsystems. Giving 
the fact that gravitational coupling is much smaller than that of electromagnetic interaction, 
coherent length of gravitons can be much longer and extends cosmological distances defined 
according to the effective metric (\ref{effmetric}). Indeed, as a non-Abelian gauge field, 
gravitons interact with each others and their condensate can have a much larger coherent length 
in the parameter space $\Xi$ than photons. In Sec. \ref{sec:hilbertfragment} we will show that 
in a classical view this process leads to the accelerating expansion of the effective geometry. 
In a quantum view, it corresponds to ergodicity and fast covering of the full parameter space 
by $\suinf$ gauge field. 

In the constraint (\ref{lagrangiangcond}) (or equivalently (\ref{enermomcond})) this $\suinf$ 
field component can be separated. In the classical limit it becomes a constant term in the 
effective Lagrangian (\ref{classicgr}) and plays the role of a cosmological constant. In 
addition, due to the global entanglement and local interaction of gravitons with matter, there 
must be a correlation between their condensate and variation of quantum state of matter, 
see Sec. \ref{sec:hilbertfragment} for quantification of this correlation. Thus, what DESI is 
observing as variation of the equation of state of dark energy with redshift $w (z)$ can be 
the feedback of the variation of both gravitational state and internal interaction of matter 
on the effective metric. The main issue of this explanation for the origin of dark energy is 
how to discriminate it from other dynamical models. \label{statfield}

\item{\bf A $\Theta$ vacuum:} This is a topologically non-trivial configuration of the 
$\suinf$ gauge field, $A_\mu^a$ at the initial time parameter $t \rightarrow -\infty$ (or its 
expectation $T\rightarrow -\infty $ as defined in Sec. \ref{app:effmetric}). We remind that 
what we call infinite past corresponds to when the Hilbert space of the Universe has 
fragmented to approximately isolated subsystems, see sections \ref{app:subsys}, 
\ref{app:mbnonthermal} and references therein for more details. A topological configuration 
on the initial 3D hypersurface of the parameter space can be included in the effective 
Lagrangian (\ref{yminvarsub}) as a 3D Chern-Simon term. It is topological and add only a phase 
to the partition function - the path integral. It can be also written as a cosmological 
constant term $\int dx^4 \sqrt{|\upeta|} \Lambda \equiv V^{(4)} \Lambda$, where $\Lambda$ is a 
constant and $V^{(4)}$ is volume of the parameter space $\Xi$ defined in Sec 
\ref{app:evolsubsys}. In the classical limit this term plays the role of the Cosmological 
Constant. However, this argument is valid only if $V^{(4)} \Lambda$ is finite. \label{topo}

\item {\bf Effective manifestation of the global entanglement:} Currently, most of the contents 
of the Universe behave classically at macroscopic scales, and quantum effects dominate only at 
short distances and time. This observation can be explained by decoherence of quantum state of 
macroscopic entities (subsystems) through their interactions with the infinite environment 
formed by other subsystems. Decoherence generates entropy and release energy. Inversely, 
entanglement needs energy~\cite{qmentangleener,qminfothermo} (and references therein). Thus, 
the global entanglement in \sqgr, reviewed in Sec. \ref{app:subsys}, needs energy. Because of 
their opposite effect on energy and entropy, it is possible that decoherence and global 
entanglement arrive to an equilibrium and result be perceived as an irreducible constant 
energy. Specifically, in (\ref{enermomequilib}) the release of energy by decohered matter is 
compensated by reduction of gravitational - $\suinf$ - energy, and perceived as dark energy 
with negative pressure. \label{entang}
\end{enumerate}

We should emphasize that the above phenomena are not mutually exclusive and all of them can 
contribute in what we observe as dark energy. Assuming only one of these processes as dominant 
contributor in dark energy, the model \ref{topo} is ruled out, if future observations confirm 
a dynamical - redshift dependent - dark energy. Moreover, its dependence on the volume of the 
parameter space or effective volume of spacetime is somehow ambiguous. Specifically, if 
$V^{(4)}$ is infinite, the contribution of Chern-Simon term should go to zero, which is 
equivalent to have a trivial, rather than topological vacuum. By contrast, models 
\ref{statfield} and \ref{entang} can accommodate a dynamical density for the effective energy 
perceived as dark energy. In addition, these processes are somehow related. Decoherence through 
interaction changes the gravitational background field and its zero mode, that may play the 
role of dark energy. Future works should include quantitative investigation of these dark 
energy candidates. They will need a better theoretical understanding of quantum aspects of 
early Universe, especially the emergence of local symmetries and subsystems, and their effects 
on the evolution of quantum gravity field $F_{\mu\nu}$\footnote{Due to the global entanglement 
and energy-momentum constraint any variation in matter sector is reflected in the $\suinf$ 
gravity sector. In particular, a species with a close to static state, for instance a 
quintessence field should generate a static gravity field $F_{\mu\nu}$. It is evident that this 
logical requirement is fulfilled by dynamic equations of fields obtained from the effective 
Lagrangian (\ref{yminvarsub}). By contrast, in general relativity, the metric is considered 
to be the fundamental gravity field. However, a field behaving similar to a cosmological 
constant leads to an approximately de Sitter geometry, which its metric is very far from being 
static. By contrast, as shown in Sec. \ref{sec:hilbertfragment}, in \sqgr~a de Sitter effective 
metric (\ref{effmetric}) corresponds to a fast variation of the quantum state. Therefore, the 
behaviour of quantum state of the Universe and effective geometry are logically consistent with 
each other.}.

According to the effective metric (\ref{effmetric}) obtained in Sec. \ref{app:effmetric} 
decoherence and emergence of approximately isolated subsystem affects the geometry of the 
effective classical spacetime (\ref{effmetric}). We describe this process in more details in 
the next section. They may have important role in early Universe inflation and late time 
accelerated expansion.

\subsection{Effect of Hilbert Space Fragmentation on the Classical Geometry}  \label{sec:hilbertfragment}
In this section we use the expression (\ref{dsave}) for the average path of the state of 
subsystems in their Hilbert space and its associated effective metric $g_{\mu\nu}$ defined in 
(\ref{effmetric}) to investigate how the emergence of subsystems and their quantum states affect 
them. 

In Sec. \ref{app:effmetric} we showed that in \sqgr~there is no standalone spacetime. 
The only shared quantity with the geometrical view of general relativity is the average or effective 
affine separation obtained in (\ref{dsave}) from quantum uncertainty relations. For this reason 
we rather investigate how the quantum state of subsystems affects the affine separation and 
through it the perceived classical cosmology. For this purpose we consider two extreme cases. 
First we assume a minimum division of quantum state of the Universe to two subsystems. In order to 
define a relative dynamics, one of them is considered as a quantum clock. In addition, we use it 
as reference subsystem for measuring area/distance scale. The other subsystem, consisting of 
the rest of the Universe, is treated as an inseparable environment. In the second case we assume 
that quantum state of the Universe is fragmented to many approximately isolated subsystems. 
As explained in Sec. \ref{app:effmetric} we concentrate on continuous parameters that 
characterize quantum state of subsystems and trace out the contribution of local symmetries.

\subsubsection{Division of the Universe to Two Subsystems}  \label{sec:2subsys}
In the first example due to the global entanglement quantum states of the clock and 
environment are entangled and inseparable. Hence, they closely follow each other - in a 
classical view their states are synchronized with each other. Specifically, if the observable 
related to the time parameter $t$ of the clock is measured, that is $t = T$ and 
$\hvarrho_c = \hvarrho_c (T)$, the state of the environment $\hvarrho_{env}$ becomes POVM 
measured: $\hvarrho_{env} = \hvarrho_{env}|_{\hvarrho_c (T, \vec{x_c})} \equiv \hvarrho_{env} (T)$. 
Moreover, as the clock subsystem is also the reference for area/distance scale, and the two 
$\suinf$ continuous parameters need a reference, without loss of generality we can assume a fixed 
value for $x_c^i,~ i=1,2,3$ components of the clock's parameters. This is equivalent to choosing a 
"rest frame" for the clock. For this reason we drop $\vec{x_c}$ from $\hvarrho_c$. Following these 
choices, the affine separation (\ref{dsave}) for this division of the Universe becomes: 
\be
ds^2 = \kappa^2 \biggl (\tr (\hvarrho_{env} d\hvarrho_{env}) \biggr )^2 \geqslant 0  
\label{2syssepar}
\ee

The last inequality in (\ref{2syssepar}) is concluded from hermiticity of $\hvarrho_{env}$ and 
$d\hvarrho_{env}$. The equality $ds = 0$ corresponds to the case where variation of state 
$d\hvarrho_{env}$ is orthogonal to the state $\hvarrho_{env}$, and thereby quantum distinguishable 
from it~\cite{qmspeed,qmspeedrev}. On the other hand, according to the classical effective metric 
(\ref{effmetric}) $ds = 0$ corresponds to events propagating on the light cone, such as 
propagation of massless particles (subsystems) or a classical Killing horizon generated by a null 
Killing vector. Thus, in \sqgr~the expression (\ref{2syssepar}) provides a quantum definition for 
the classical concept of horizon. 

According to the Mandelstam-Tamm inequality and its extension to Hilbert space geometry 
(\ref{ineqgen}), states that violate them, that is when $ds^2 < 0$, are indistinguishable due to 
the quantum uncertainties. Consequently, a black hole can be defined as a fully entangled and 
inseparable quantum system and its semi-classical Hawking evaporation as excitations that reduce 
or breaks this entanglement.

Using the Cauchy-Schwarz inequality 
$(\tr \hA^\dagger \hB)^2 \leqslant \tr(\hA^\dagger \hA) \tr(\hB^\dagger \hB)$ and properties of 
density matrices, we find:
\be
ds^2 \leqslant \kappa^2 \tr (\hvarrho_{env}^2(T)) ~ \tr (d\hvarrho_{env}^2(T)) \leqslant \kappa^2 
\tr (d\hvarrho_{env}^2(T))  \label{2sysseparineq}
\ee
The above inequalities show that in the perceived classical geometry, the affine separation 
between events is a lower limit for the effective - traced - variation of cosmological quantum 
state. Finally, the relation between $ds$ and quantum state in (\ref{dsave}) is based on the 
relative purity $F$, defined for state $\hvarrho (t_0)$ and its evolution $\hvarrho (t_1)$ as:
\be
F \equiv \frac{\tr (\hvarrho (t_0) \hvarrho (t_1))}{\tr(\hvarrho^2 (t_0))}  \label{relpurity}
\ee

Thus, the affine separation in the Hilbert space $ds' \equiv \kappa^{-1} ds$ can be related to 
purity change:
\bea
F + dF &=& \frac{\tr (\hvarrho_{env} (\hvarrho_{env} + d\hvarrho_{env}))}{\tr \hvarrho_{env}^2}  
\label{fidelityvar} \\
ds' &=& \sqrt{\tr \hvarrho_{env}^2} dF = \tr (\hvarrho_{env} d\hvarrho_{env})   \label{affinevsfidelity}
\eea

From definition of relative fidelity (\ref{relpurity}) it is evident that 
$0 \leqslant F \leqslant 1$. Thus, irrespective of how the state is evolved with time, the affine 
separation $ds'$ is bounded. For example, assume that at a given time $T$ the 
state $\hvarrho_{env}$ is a completely incoherent state, that is all off-diagonal components of 
$\hvarrho_{env}$ are zero and diagonal components $P (T, \vec{x})$ are normalized to 
$\int d^3 x \sqrt{|\upeta^{(3)}|} P (T, \vec{x}) = 1$. Assuming an approximately smooth spectrum 
for $P (T, \vec{x})$, the non-zero diagonal components of $\hvarrho_{env}$ are of order of their 
average $\bar{P} (T) \sim 1/V$, where $V$ is the volume of the 3D parameter space at time $T$. 
In this case variation of diagonal components is of order 
$d\bar{P} (T)$. From this estimation and the second identity in (\ref{affinevsfidelity}) we 
conclude that $ds' \sim d\bar{P} (T) \ll \bar{P} (T)$. From the first identity we obtain: 
\be
dF \sim d\bar{P} / \sqrt{V} \sim ds / (\kappa \sqrt{V})  \label{fidelityaffine}
\ee
Thus, if the quantum state of the Universe has a smooth spectrum, variation of the classical 
affine separation reflects the variation of the quantum state. This case corresponds to a well 
spatially mixed Universe. In the mode space this corresponds to the dominance of one mode and 
approximate thermalization of eigen states. 

\subsubsection{Division of the Universe to Many Subsystems}  \label{sec:manysubsys}
The second example we consider here is a Universe highly fragmented to approximately separable 
subsystems. Similar to the previous example, one of them can be considered as a clock. Moreover, 
in presence of many subsystems the area/distance reference and the clock can be different, and 
the latter is not in general entangled to any single subsystem, but due to 
the global entanglement, it is only entangled to their ensemble, which is the environment for 
the clock.

According to the above assumptions the environment $\hvarrho_{env}$ is in a product state (Product 
of states are assumed to be symmetrized): 
\be
\hvarrho_{env}  \approx \otimes_i \hvarrho_i  \label{envstateprod}
\ee

We use the product form of $\hvarrho_{env}$ and assume that $||d\hvarrho_i|| \ll ||\hvarrho_i||$. 
Thus, at lowest order in $||d\hvarrho_i||~\forall i$, the variation of $\hvarrho_{env}$ can be 
written as:
\be
d\hvarrho_{env} \approx \sum_i \hvarrho_1 \otimes \hvarrho_2 \ldots \otimes d\hvarrho_i 
\otimes \hvarrho_{i+1} \otimes \ldots  \label{drhoenv}
\ee

Inserting this expression and (\ref{envstateprod}) in (\ref{2syssepar}) leads to:
\be
ds'^2 \approx \sum_i \tr (\hvarrho_1^2 (T)) \tr (\hvarrho_2^2 (T)) \ldots \tr (\hvarrho_i (T) 
d\hvarrho_i (T)) \tr (\hvarrho_{i+1}^2(T)) \ldots \leqslant \sum_i \tr (\hvarrho_i (T) 
d\hvarrho_i (T)) \label{ds2drhoenv}
\ee

The last inequality is due to $\tr (\hvarrho_i^2) \leqslant 1, ~ \forall i$. Comparison of 
(\ref{ds2drhoenv}) with (\ref{2sysseparineq}) shows that the r.h.s. of (\ref{ds2drhoenv}) is 
approximately proportional to the number of separable subsystems and variation of their states. 
For instance, assume completely incoherent states for separable subsystems. Such quantum 
mixed states correspond to classical statistical states. For weak interaction and entanglement 
among subsystems and environment $\tr \hvarrho_i^2 \lesssim 1$. Thus, if we apply the order of 
magnitude estimation method discussed in Sec. \ref{sec:2subsys} to (\ref{ds2drhoenv}), we find:
\be
ds2 \equiv g_{\mu\nu}(X) dX^\mu dX^\nu  \sim \kappa^2 N \llangle dP \rrangle   \label{dsapprox}
\ee

The double bracket represent averaging over subsystems and their distribution of states. 
As explained in Sec. \ref{app:classiclimit}, the exact description of $g_{\mu\nu}(X)$ should be 
calculated from classical limit of \sqgr. Nonetheless, here our purpose is to estimate its general 
properties directly from quantum state of the Universe\footnote{In this section the expression 
{\it quantum state of the Universe} means the state of its contents - its subsystems.}. The 
reason is the fact that at present the behaviour of the effective metric at different 
cosmological epochs is the only accessible measurable of the quantum state of the Universe. 
For this purpose we write the effective metric $g_{\mu\nu}(X)$ in conformal gauge, which is more 
suitable for assessing the effect of quantum state. Moreover, we neglect its spatial dependence. 
Under these assumptions and simplifications, we find:
\be
ds2 \equiv a^2(\tau) (d\tau^2 - \delta_{ij} dX^i dX^j) \sim \kappa^2 N \llangle dP \rrangle 
\label{dsconformapprox}
\ee
where $\tau$ is the conformal time. 

Observations show that the expansion factor $a (\tau)$ is a monotonously increasing function of 
time. In particular, during inflation, which seems necessary for explaining present state of the 
Universe, according to the classical general relativity the effective geometry is approximately 
de Sitter with $a(\tau) = -1/H\tau, \tau = [-\infty, 0]$. For a fixed coordinate separation 
$\Delta_s \equiv d\tau^2 - \delta_{ij} dX^i dX^j$ at late times the affine separation diverges, 
that is $ds \rightarrow \infty$ for $\tau \rightarrow 0$. If we assume finite variation of state 
$\llangle dP \rrangle$ for subsystems, according to (\ref{dsconformapprox}), their number should 
diverge - or more realistically grows exponentially, similar to what is observed in the Hilbert 
space fragmentation of many-body systems~\cite{qmhilbertspacefragex}. This result shows that in the 
framework of \sqgr~inflation and particle production can happen at the same time. Moreover, what 
is called inflaton field may be an order parameter, rather than a fundamental field. Indeed, 
considering the early Universe as a many-body quantum system,  emergence 
of exotic many-body phenomena is expected. Similar to condensed matter, these phenomena 
would depend on non-local orders and symmetries that are best understood through order 
parameters and effective fields. A more quantitative study of these processes 
need a better theoretical understanding of the Hilbert space fragmentation, which is out of the 
scope present work. 

More generally, the r.h.s. of (\ref{dsconformapprox}) shows that the affine separation 
$ds$ grows with the number of approximately isolated subsystems. Therefore, the apparent 
expansion of the Universe can be interpreted as growth of the number of approximately isolated 
quantum subsystems. Of course, variation of their quantum states, approximated by 
$\llangle dP \rrangle$ in (\ref{dsapprox}) also affect the expansion factor. Indeed, as 
$g_{\mu\nu}(X)$ and its associated classical spacetime present the average path of the quantum 
state in its parameter space $\Xi$, the perceived expansion of the Universe can be interpreted as 
the tendency of quantum state to approach ergodicity - thermalization. 
Specifically, according to the Eigenstate Thermalization Hypothesis (ETH), briefly reviewed in 
Sec. \ref{app:mbnonthermal}, it is expected that after the epoch of fast fragmentation, some of the 
components in the r.h.s. of (\ref{envstateprod}), and thereby (\ref{ds2drhoenv}), evolve to a 
thermal state. This is consistent with present cosmological observations. However, as the available 
parameter space $\Xi$ is infinite and ergodicity requires filling the whole available parameter space, 
thermal states should cool and dissipate. This conclusion of the model is also consistent with 
cosmological observations. On the other hand, according to the axioms of \sqgr~the Universe can 
never be {\it empty}. Therefore, some of the components of the product state (\ref{envstateprod}) 
should not thermalize or be in a dissipative state. As explained in Sec. \ref{sec:deoldmodels}, 
one such candidate subsystem is the condensation of spin-1 gravitons, which in \sqgr~are 
omnipresent. This component (subsystem) can be formulated as an extended (or generalized) 
Glauber coherent state $|\Psi_{GC} \rangle$~\cite{coherestate,houricond}:
\be
|\Psi_{GC} \rangle = \sumint A_k e^{C_k a^\dagger_k} |\emptyset\rangle   \label{gencoherstate}
\ee
where $k$ collectively presents all the parameters characterizing one particle states 
$\Phi (k) \equiv a^\dagger_k |\emptyset \rangle$. Although $|\Psi_{GC} \rangle$ includes 
superposition of states with infinite number of elementary field $\Phi (k)$, only its component
condensate with a fixed $k$ can be considered as an approximately isolated subsystem, and a 
factor in the tensor product (\ref{envstateprod}). For spin-1 gravitons in \sqgr~parameters $k$ 
are continuous. Thus, the condensate state consists of infinite number of components, and 
in analogy with the stochastic gravitational waves, the condensate direction as a vector field of 
the parameter space $\Xi$ would be random. Moreover, the graviton 
condensate interacts with itself and other contents of the Universe through $\suinf$ symmetry - 
the gravity and is correlated and entangled to them. Thus, as states of other subsystems in 
(\ref{envstateprod}) change with time, so does $|\Psi_{GC} \rangle$, but more slowly - its 
variation is dumped due to the presence of very large number of entangled particles. If the 
graviton condensate is the dominant contributor to dark energy, its slow dynamics seems 
consistent with the claim of a dynamical dark energy after recent cosmological observations. 
As mentioned earlier a more quantitative study of these process needs 
a better understanding of HSF and formation and evolution of graviton condensate, which 
we leave to future works. 

The approximation (\ref{dsconformapprox}) cannot be used to estimate the effect of a generalized 
coherent state $|\Psi_{GC} \rangle$ on the effective metric. Nonetheless, there are infinite number 
of condensated modes indexed by $k$, which includes coordinates $(\eta, \zeta)$ of the 
diffeo-surface of $\suinf$ symmetry. Therefore, if graviton condensate becomes the only content 
of the Universe, its effect on the effective geometry would be similar to the effect of HSF. 
Accordingly, the corresponding effective metric becomes approximately de Sitter. However, as 
discussed in Sec. \ref{app:sqgrrev} and with more details in~\cite{hourisqgr, hourisqgrym}, 
gravitational interaction is much weaker than those of internal symmetries $G$. Moreover, 
dissipation of subsystems distinguishable by their internal symmetries leads to loss of the 
notion of quantum clock, time, and quantum state of the Universe becomes once again 
topological (reviewed in Sec. \ref{app:evoluniv}). And a new evolution cycle of the Universe arises. 

\section{Conclusion}  \label{sec:outlines}
In this work we investigated how the classical concept of energy-momentum arises in \sqgr, in which 
spacetime is not a fundamental entity but emergent. We showed that independence of the Lagrangian 
from arbitrary geometry of the parameter space of the Hilbert space leads to an equation similar 
to the Einstein equation in classical gravity. Thus, in the framework of \sqgr~it is a constraint 
rather than equation of motion for gravitational field. Accordingly, the Einstein-Hilbert action 
loses its usual interpretation, except in classical limit, where it replaces $\suinf$ gauge term 
in the action and becomes the geometric part of the Einstein equation. We have also defined 
Hamiltonian and Lagrangian superoperators. They can be directly applied to quantum state of the 
Universe and its fields/subsystems to study their evolution. 

The constraint equation includes terms analogous to energy-momentum tensor in general relativity 
and QFT. In particular, at lowest quantum order, one of these terms has the form of energy-momentum 
tensor for the $\suinf$ Yang-Mills gauge field and can be interpreted as energy-momentum of 
gravity. This concept does not exist in general gravity, because the usual definition of 
energy-momentum tensor of matter coincides with how the Einstein equation is obtained from 
Einstein-Hilbert action. Therefore, \sqgr~put gravity in the same footpath of other contents of 
the Universe. In addition, the energy-momentum constraint can be interpreted as a null total 
energy-momentum for the Universe. This makes physical sense, because energy-momentum is related 
to Poincar\'e symmetry. But, there is nothing outside the Universe such that, for instance it moves 
or rotates with respect to it. Thus, the total energy-momentum of the Universe must be null. 
Interestingly, it can be shown that in any reparameterization invariant QGR 
model in which the classical spacetime and its metric are not fundamental but emergent, 
the Einstein equation arises as a constraint (work in preparation). 

In \sqgr~there is no clear candidate for dark energy. Moreover, we showed that due to 
the absence of a standalone spacetime the usual concept of vacuum does not exist. Nonetheless, 
many of dark energy candidates in the literature and a few \sqgr~specific models can explain 
the observed close to constant term in the classical Einstein equation. In addition, we used the 
relationship between quantum state of the contents of the Universe, affine separation of effective 
metric, and their variations to evaluate their effect on the apparent expansion of the Universe. 
In particular, we discussed conditions for emergence of spin-1 graviton condensate that may 
play an important role in the early Universe inflation and late time accelerating expansion. For 
this purpose, we quantified variation of the state through measures of its coherence and also 
overlap between states. These measures are also useful for describing the process of clustering 
and fragmentation of the quantum state of the Universe - in other words the Hilbert space 
fragmentation of the Universe to approximately isolated subsystems. These measures may also have 
more general application in the study of the Hilbert space fragmentation in condensed matter 
models.

Candidate processes for dark energy and their consequences for the currently observable 
cosmological features need more detailed study, including a better understanding and 
quantification of the Hilbert space fragmentation. This is not an easy tasks, because even 
for laboratory testable condensed matter models dynamics of fragmentation is not well understood. 
More information about particle physics of the early Universe and/or high energy processes may 
help to understand how fragmentation locally have reduced symmetries of the contents (subsystems) 
of the Universe to the smallest rank non-factorizable non-Abelian groups, namely 
$SU(2) \times U(1)$ and $SU(3)$ of the Standard Model of particle physics at low energies.

\appendix
\section{$\boldsymbol{SU(\infty)}$-QGR~in a Nutshell}\label{app:sqgrrev}
As \sqgr~is a new model, in this section we highlight its main properties, specially those used 
in the main part of this work. Readers familiar with the model may skip this section.

\subsection{Axioms and Hilbert Space}  \label{app:axiom}
The \sqgr~is an axiomatic and abstract approach to quantum cosmology 
and gravity. Motivations and rationale for its construction are presented 
in~\cite{houriqmsymmgr,hourisqgr}. It is reviewed in~\cite{hourisqgrhighlight}. 
Axiom of \sqgr~can be summarized as the followings: (1) Quantum mechanics applies at all 
scales, including the Universe as a whole; (2) Hilbert spaces of quantum systems are 
exclusively determined by their symmetries and represent them; (3) There are infinite number 
of mutually commuting observables in the Universe. The last axiom means that the Hilbert 
space of the Universe $\hm_U$ represents $\suinf$ symmetry group\footnote{We could 
consider $U(\infty) = \suinf \times U(1)$ as symmetry represented by $\hm_U$. However, as 
the Hilbert space of quantum systems are projective, the corresponding Hilbert space would 
be the same.}~\cite{suninfhoppthesis,suninfvirasoro,suninfvirasoro0,suninfsurfaceanomal,suinftriang,suninftorus,suinftriang0,adifftorussuinf,suninfrep,suninfsimplect,suninfrep0}, 
see e.g. appendices of~\cite{houriqmsymmgr,hourisqgrym} for review of $\suinf$ and its 
representations. There is no background spacetime in this model and the usual quantization 
procedure is replaced by non-commutative $\suinfa$ algebra of linear operators 
$\hO \in \bm[\hm_U]$ acting on $\hm_U$~\cite{qgrnoncommut,qmmathbook}. Specifically, it has 
been proven that $\suinf$ is homomorphic to the group of area preserving diffeomorphism of 
2D Riemann compact surfaces $D_2$, that is $\bm[\hm_U] \cong SU(\infty) \cong ADiff(D_2)$ 
\footnote{In this work the symbol $\cong$ indicates homomorphism}. We call these surfaces 
{\it diffeo-surface}. The $\suinfa$ algebra is homomorphic to Poisson bracket of functions defined 
on a diffeo-surface $D_2$: 
\be
\hL_f = \frac{\partial f}{\partial \eta} \frac{\partial }{\partial \zeta} - 
\frac{\partial f}{\partial \zeta} \frac{\partial }{\partial \eta} \quad , 
\quad \hL_f ~ g = \{f,g\} \label{suinfgendef}
\ee
where $f$ is any $C^\infty$ scalar function on $D_2$. The Lie algebra of these operators is: 
\be
[\hL_f, \hL_g] = \hL_{\{f,g\}},  \label{suinfal}
\ee

Using an orthonormal function basis, for instance spherical harmonic functions $Y_{lm}$, 
generators of $\suinfa$ algebra can be explicitly described with respect to coordinates of 
$D_2$:
\bea
\hL_{lm} (\theta, \phi) & = & i\hbar ~ \biggl (\frac{\partial Y_{lm}}{\partial \cos \theta} 
\frac{\partial}{\partial \phi} - 
\frac{\partial Y_{lm}}{\partial \phi} \frac{\partial}{\partial \cos \theta} \biggr ), 
\quad \quad [\hL_{lm} , \hL_{l'm'}] = -i\hbar f ^{l"m"}_{lm,l'm'} \hL_{l''m''}
\nonumber \\ 
&& {\mu,~\nu \in \{\theta,\phi \}}, \quad l\geqslant 0,~ -l \leqslant m \leqslant l 
\label{lharminicexp}\\
\hL_{lm} Y_{l'm'} & = & -i\hbar \{Y_{lm},~Y_{l'm'}\} = -i\hbar f ^{l"m"}_{lm,l'm'} Y_{l"m"}, 
\label{lapp}
\eea
and structure coefficients $f ^{l"m"}_{lm,l'm'}$ of $\suinfa$ can be calculated with respect to 
3j symbols~\cite{suninfhoppthesis}. The second equality in (\ref{lapp}) shows that spherical 
harmonic functions can be considered as generators $ADiff \cong \suinf$. Although the pair 
$(l,m)$ that characterize generators of $\suinfa$ are discrete, the expansion to 
spherical harmonics can be inverted - at least formally - and generators can be expressed with 
respect to coordinates $(\theta,\phi)$ of the diffeo-surface $D_2$ only~\cite{houriqmsymmgr}. 
Therefore, the symmetry group $\suinf$ and its algebra are characterized by two continuous 
variables and vectors of the Hilbert space are complex valued functions depending on them.

\subsubsection{Quantization}  \label{app:qm}
The Planck constant $\hbar$ is explicitly shown in the description of $\suinf$ generators and 
its algebra to remind that in analogy with non-commutative geometry 
models~\cite{qgrnoncommut,qgrmatrixnoncommut} they replace the standard Heisenberg 
uncertainty relations used for quantization of classical models. Nonetheless, due to the fact 
that $\suinfa$ generators are characterized by two continuous parameters, the standard 
Heisenberg uncertainty relations exist between $(\theta, \phi)$ - considered as operators - 
and their duals~\cite{houriqmsymmgr,hourisqgr}. Nonetheless, they can be expanded with 
respect to the generators $\hL_a$ of $\suinfa$, but their description is not unique. 
This means that there is much more information in $\hL_a$ than in characterizing parameters 
and their duals.

If $\hbar \rightarrow 0$, the algebra (\ref{lharminicexp}) becomes Abelian, and the structure 
of the model - the Universe - collapses because there is no background space time. Hence, 
\sqgr~is inherently quantum and does not have an asymptotic "classical" limit in the usual 
sense.

\subsection{Dynamics of the Whole Universe}  \label{app:evoluniv}
As expected, in absence of a background spacetime the above Universe is static. Nonetheless, 
a $\suinf$ invariant functional Lagrangian $\lm_U$ can be defined by using traces of 
products of the $\suinf$ generators $\hL_{lm}$. After imposing the logical constraint that 
this Lagrangian should be invariant under coordinate transformation of the diffeo-surface, 
it is shown~\cite{hourisqgrcomp,hourisqgr} that at lowest order in the number of $\suinf$ 
generators in the traces $\lm_U$ takes the form of a $\suinf$ Yang-Mills model defined on the 
2D diffeo-surface:
\vspace{-9pt}
\bea
\lm_U & = &\int d^2\Omega ~\lm_U (\theta,\phi) = \int d^2\Omega ~ \biggl [ ~\frac{1}{2} ~ 
\tr (F^{\mu\nu} F_{\mu\nu}) + \frac {1}{2} \tr (\sD \hvarrho_U) \biggr ], \quad \quad 
d^2\Omega \equiv d^2x \sqrt{|\upeta^{(2)}|} \nonumber \\
&& \label{lagrange2d} \\
F_{\mu\nu} & \equiv & F_{\mu\nu}^a \hL^a \equiv [D_\mu, D_\nu], \quad 
D_\mu = (\partial_\mu - \Gamma_\mu) + \sum_a i \lambda A_\mu^a (x) \hL^a (x), \mu,~\nu = 0,~1 
\label{yminvardef}
\eea
where $\upeta^{(2)}$ is the determinant of the 2D metric of the diffeo-surface and $\Gamma_\mu$
is the corresponding connection. Expression of covariant derivative $\sD$ depends on the 
representation of Euclidean symmetry of the 2D parameter space by the density operator 
$\hvarrho_U$, see~\cite{hourisqgr} for examples. The index $a$ indicates parameters of the 
orthogonal basis used for the description of generators of $\suinf$ symmetry. For instance, 
in spherical basis (\ref{lharminicexp}) $a=(l,m)$; in torus 
basis~\cite{suinftriang,suninftorus,suinftriang0,suninfrep} (reviewed in~\cite{hourisqgrym}) 
it is a 2D vector with integer components. 

It is well known that 2D pure Yang-Mills Lagrangian, that is the first term in 
(\ref{lagrange2d}) is topological. Moreover, in the case of $\lm_U$ any modification of 
the metric $\upeta_{\mu\nu}$, that is a diffeomorphism of the diffeo-surface $D_2$ corresponds 
to a $\suinf$ gauge transformation, under which the Lagrangian $\lm_U$ is invariant 
up to an irrelevant and unobservable global scaling of $D_2$. Moreover, 
$\hvarrho_U \in \bm[\hm_U] \cong \suinf$. Thus, this term is also independent of the geometry 
of $D_2$ up to an irrelevant scaling. Consequently, (\ref{lagrange2d}) is a topological field 
theory. Giving the fact that 2D manifolds have only one topological class, namely the Euler 
characteristics, the integrand of (\ref{lagrange2d}) is proportional to the scalar curvature of 
$D_2$ and $\lm_U$ is proportional to the Euler constant, uniquely defined by the genus of $D_2$. 
These properties establish the relationship between the abstract quantum system constructed 
under axioms described in Sec. \ref{app:axiom}, and geometric and topological properties of 
the associated 2D diffeo-surface. 

It is demonstrated~\cite{houriqmsymmgr} that the application of variational principle to 
$\lm_U$ leads to static and trivial solutions for amplitudes $L_a$. Moreover, according to 
the axioms, any state can be transformed to another one by a $\suinf$ transformation, under 
which observables are conserved. Another important conclusion from these properties is that 
amplitudes $L_a$'s cannot be all zero, because the zero vector is always projected to 
itself. The same conclusion also applies to components of the density operator $\hrho_U$. 
Thus, a zero state for the Universe means no Universe. In other words, in the framework of 
\sqgr~an empty - vacuum - Universe does not exist. This is in contrast to classical or quantum 
models, in which spacetime is treated as a standalone physical entity that its existence does 
not depend on the presence of its contents.

\subsection{Fragmentation to Subsystems} \label{app:subsys}
Despite global triviality of the Universe, quantum fluctuations, that is 
random application of operators $\hO \in \bm[\hm_U]$, leads to clustering and approximate 
factorization of the Hilbert space: 
\be
\hm_U \cong \bigotimes_{i=1}^{\infty} \hm_i  \label{hilbertfactor} 
\ee
This means an approximate division of the Universe to subsystems according to the criteria 
defined in~\cite{sysdiv}.

The process of Hilbert space fragmentation of closed quantum systems to subsystems with their 
own symmetries $G_i$ is observed in many condensed matter models, 
see e.g.~\cite{qmhilbertspacefragex,qmfragalgebra,qmfragsimul} and reviews~\cite{qmfragrev,qmfragrev0}. 
In Sec. \ref{app:mbnonthermal} we review the emergence of localized and clustered states in 
isolated many-body quantum systems and how these observations can be relevant for the structure 
of the Universe in the framework of \sqgr. In addition, it is proved~\cite{hourisqgr,hourisqgrym} 
(and references therein) that:
\be
\suinf \cong \otimes_{i=1}^\infty G_i  \label{suinffactor}
\ee
Therefore, it is not unreasonable to assume that $\hm_U$ becomes a product of Hilbert spaces 
similar to (\ref{suinffactor}), consisting of approximately isolated subsystems. In addition to 
local interaction convoyed by $G_i$ symmetries, they interact would interact through the universal 
$\suinf$ symmetry. Thus, the only truly isolated quantum system is the whole Universe.

Specifically, according to the second axiom of the model and criteria for division of a quantum 
system to subsystems~\cite{sysdiv}, Hilbert spaces of subsystems represent a reduced subalgebra 
and its associated finite rank, group generically called $G$, see~\cite{hourisqgr,hourisqgrym} 
for details. Nonetheless, in \sqgr~subsystems are only approximately isolated and irrespective of 
their local symmetries $G$, they interact through the global $\suinf$ symmetry with an environment 
generated by the rest of the Universe and are entangled to it. We call this property 
``the global entanglement'', see~\cite{hourisqgr,hourisqgrym} for the proof and more explanation. 
Therefore, the full symmetry of each subsystem is $G \times SU(N\rightarrow \infty)$, but symmetry 
of the whole Universe remains unchanged, because $\suinf^n \cong \suinf \forall n$. Note that 
the concept of subsystem here is more general than fundamental fields in QFT. Those subsystems 
can be defined as special cases where the symmetry $G$ is irreducibly represented and the 
subsystem cannot be divided further. 

\subsection{Parameter space of subsystems}  \label{app:paramsubsys}
For a single representation of $\suinf$ group the area of the corresponding diffeo-surface is 
irrelevant. Indeed, \ref{suinfal}) and (\ref{lharminicexp}) generates the algebra 
$\suinfa \times \um(1)$~\cite{suninfrep}. However, in presence of multiple representations, the 
relative areas of diffeo-surfaces (or equivalently the relative phases of their $\um(1)$) 
become observable. Moreover, one of the subsystems can be considered as a clock and one 
of its observables as a time parameter. Thus, a relative dynamics can be defined for subsystems 
\`a la Page and Wootters~\cite{qmtimepage} or equivalent methods~\cite{qmtimedef}. 

After choosing a reference subsystem for area (or size/distance) scale and a clock, states of 
subsystems are characterized by parameters defining their representation of their 
{\it local symmetry} $G$ and by 4 continuous parameters: two of them specify the representation 
of $\suinf$ symmetry, one determines area or size/distance scale, and the last parameter is time 
that characterizes the relative dynamics with respect to the chosen clock. Despite their distinct 
origin, these parameters are not distinguishable and states of subsystems are in general in their 
superposition. In this work we call this 4D parameter space $\Xi$. Quantum states, operators, and 
processes must be invariant under diffeomorphism of $\Xi$.

Because of the global entanglement, in order to define the quantum state of one subsystem in 
isolation we must trace out the contribution of the rest of the Universe. The resulting density 
matrix $\hvarrho_G$ takes the following form~\cite{hourisqgrym}:
\be
\hvarrho_G \equiv \tr_\infty \hvarrho_U = \sum_{\substack{\{k_G, k'_G\}, \\ \{y\}}} 
A_G (k_G; y) A^*_G (k'_G, y) ~ \hrho_G (k_G , k'_G), \quad \quad y \equiv (t, r,\theta, \phi) 
\label{gdensity}
\ee
where $\hvarrho_U$ is the state of the Universe as a whole; $\{k_G\}$ is the set of possible 
values for parameters of the representation of $G$ symmetry; and $\hrho_G (k_G , k'_G)$ is a 
basis for $\bm [\hm_G] \cong G$. It is straightforward to show that $\hvarrho_G$ is a mixed 
state~\cite{hourisqgrym}. This feature is reflected in the dependence of amplitudes 
$A_G (k_G; y)$ on the $\suinf$ related parameters $y$, which play the role of a `{\it classical} 
background. Similarly, we can trace out the contribution of local symmetry and obtain the 
mixed state $\hvarrho_\infty$ in which $\{k_G\}$ are {\it external} parameters. These states can 
be purified:
\be
|\Psi_{G_\infty}\rangle \equiv \sum_{\{k_G\}; (\eta, \zeta, \cdots)\}} 
A_{G_\infty} (k_G; \eta, \zeta, \cdots) ~ |\psi_G (k_G) \rangle \times 
|\psi_\infty (x) \rangle  \label{gpurified}
\ee
where $|\psi_G (k_G) \rangle$ and $|\psi_\infty (x) \rangle$ are bases for $G$ and $\suinf$ 
symmetry representing Hilbert spaces, respectively. Note that although $\suinf$ representations 
depend only on two continuous parameters, due to the inseparability of $\Xi$, all 4 parameters 
appear in the basis. See also the next section for more clarification.

\subsection{Dynamics of the Subsystems}  \label{app:evolsubsys}
Similar to the case of the whole Universe, a Lagrangian functional can be constructed on the 4D 
continuous parameters space $\Xi$ of subsystems.from invariants of $\suinf$ and $G$ symmetries.  
After imposing the invariance under redefinition of parameters, that is Poincar\'e symmetry,, 
at lowest order of the number of generators in the traces the effective action takes the form of 
a non-Abelian Yang-Mills gauge model for both $\suinf$ and $G$ symmetries:
\vspace{-9pt}
\bea
&& \lm_{U_s} = \int d^4x \sqrt{|\upeta|} ~ \biggl [\frac{\kappa}{4} 
\tr (F^{\mu\nu} F_{\mu\nu}) + \frac{1}{4} \tr (G^{\mu\nu} G_{\mu\nu}) + 
\frac {1}{2} \sum_s \tr (\sD \hrho_s) \biggr ], \quad \mu,~\nu = 0, \cdots,~3  \nonumber \\
&& \label{yminvarsub} \\
&& F_{\mu\nu} \equiv F_{\mu\nu}^a \hL^a \equiv [D_\mu, D_\nu], \quad D_\mu = 
(\partial_\mu - \Gamma_\mu) \mathbbm{1}_\infty - i \lambda A_\mu, \quad A_\mu, \equiv 
\sum_a A_\mu^a \hL^a, 
\label{yminvardefsuinf} \\
&& G_{\mu\nu} \equiv G_{\mu\nu}^{a'} \hT^{a'} \equiv [D'_\mu, D'_\nu], \quad D'_\mu = 
(\partial_\mu - \Gamma_\mu - i \lambda A_\mu) \mathbbm{1}_G - i \lambda_G B_\mu, \quad
B_\mu \equiv \sum_{a'} B_\mu^{a'} \hT^{a'} 
\nonumber \\
&& \label{yminvardefg}
\eea

The first and second terms in (\ref{yminvarsub}) are the Lagrangian density for the $\suinf$ 
and internal symmetry $G$ gauge fields $A_\mu^a$ and $B_\mu^a$, respectively. The $\suinf$ 
generators $\hL^a$ and index $a$ are the sane as the case of $\lm _U$. Operators $T^{a'}$ 
are generators of the internal symmetry $G$ of the subsystem. Their number must be finite, 
because $G$ has a finite rank. The density matrix $\hrho_s$ is the quantum state of the 
subsystem and the covariant differential operator $\sD$ depends on the representation 
of symmetries of the 4D parameter space $\Xi$ by $\hrho_s$ see~\cite{hourisqgr} for details 
and examples. The constant $\kappa$ is used for the adjustment of units in the classical 
limit as defined below. It is also used for interpretation of the constraint (\ref{lagrangiangcond}).

The function $\Gamma_\mu$ is the geometric connection for the metric $\upeta_{\mu\nu}$ of 
the parameter space $\Xi$. In~\cite{hourisqgr,hourisqgrym} it is shown that $\upeta_{\mu\nu}$ 
is arbitrary. Indeed, the effect of connection $\Gamma_\mu$ can be neutralized by a $\suinf$ 
gauge transformation. Therefore, geometry of $\Xi$ is not a physical observable and despite 
its similarity with the perceived classical spacetime, $\Xi$ cannot be identified as the latter. 
Nonetheless, in Sec. \ref{sec:enermom} we investigate the role of $\upeta_{\mu\nu}$ in the 
definition of observables, in particular the effective energy-momentum tensor. The Yang-Mills 
nature of the Lagrangian (\ref{yminvarsub}) for the $\suinf$ gauge sector, which we interpret as 
quantum gravity, means that according to \sqgr~its mediator boson is a spin-1 field. 

A comment is order about the interpretation of the Lagrangian $\lm_{U_s}$. As the integrand 
and the total Lagrangian are numbers rather than operators, the question arises whether 
$\lm_{U_s}$ presents a quantized dynamics or we have to quantize it. The second
case would be in contradiction with our claim that non-commutativity of the algebra of 
operators is sufficient for considering \sqgr~inherently quantum. To clarify this ambiguity, 
we first consider the density term. For pure states it can be written as 
$|\psi\rangle \langle \psi|$, where $|\psi\rangle$ is a vector in the Hilbert space of one 
subsystem. Thus, this term is analogous to a 1-particle Lagrangian for a matter field 
with infinite number of {\it internal} states - colors.  A basis for the state vectors $|\psi\rangle$ 
has the form of a column matrix $(0, 0,...,1, 0, 0, .....)^T$, where $T$ means transpose. 
The coefficients in the expansion are amplitudes, which their norm can be interpreted as 
probability, just like any wavefunction in QM. They should not be quantized, because they 
already present a quantum state. Their dependence on the 3+1 parameter space is similar 
to QM in Minkowski background. The same interpretation applies to $\suinf$ gauge term. 
But, here gravitons are in fundamental representation of $\suinf$, described by a square 
matrix. In the same manner they can be expanded using a basis, and interpretation of each 
term is similar to that of the density term. They are analogous to W and Z in Electroweak, if its 
symmetry was not broken. As usual, we can use this 1-particle description to construct a 
Fock space. Alternatively, we may use $\lm_{U_s}$ similar to a classical Lagrangian on 
an independent background and use e.g. path integral for {\it quantization}. In this case, 
we must keep in mind that the parameter space $\Xi$ is indeed generated by the ensemble of 
subsystem, and in contrast to QFT's in curved spacetimes, there is no backreaction on the 
$\Xi$, because its contents are just indices. Comparing it with the strong interaction, the 
space (set) of 3 colors of quarks or 8 colors of gluons are always the same and independent 
of the interaction of particles to which they are associated.

\subsection{Emergence of Classical Spacetime and Geometry}  \label{app:effmetric}
It is important to emphasize that in \sqgr~there is no fundamental spacetime. Nonetheless, 
the perceived classical spacetime can be related to quantum states of the contents of the Universe  
and their parameter space $\Xi$. To establish and explain this relationship, we use the concept 
of Quantum Speed Limit (QSL) which was first introduced through Mandelstam-Tamm (MT) 
inequality~\cite{qmspeed}. The origin of QSL is the Heisenberg uncertainty relation, see 
e.g~\cite{qmspeedrev} for a comprehensive review or~\cite{hourisqgrhighlight} for a brief one. 
The MT-QSL is an attainable limit for pure state. More generally, for any quantum state $\hrho$ - 
pure or mixed - and any type of evolution - unitary, Markovian, or non-Markovian - a minimum 
(measured) time $\Delta T = T - T_0$ is necessary to change the state $\hrho_0 \equiv \hrho (T_0)$ 
of a quantum system to $\hrho \equiv \hrho (T)$. Here $T$ indicates the {\it measured time}, 
rather than time parameter $t$. Using a distance function ${\dm}$ for quantum states defined 
on their Hilbert space $\hm$ and its $\bm[\hm]$, for instance the trace distance, the QSL's can 
be written as~\cite{qmspeedrev}(and references therein):
\be
\Delta T \geqslant \frac{{\dm} (\hvarrho(T_0), \hvarrho(T))}{\llangle \sqrt{{\tt g}_{tt}} \rrangle}  
\label{ineqgen}
\ee 
where ${\tt g}_{tt}$ is the metric used for the definition of distance function $\dm$ of density 
operators and depends on the time parameter $t$. The double bracket means averaging over 
the measured time interval $\Delta T$. For an infinitesimal evolution of the density matrix,  
(\ref{ineqgen}) defines a path in $\bm[\hm]$ with a path parameter $ds$. This quantity depends 
only on the variation of the state and depends on the chosen quantum clock only through 
variation of the state $\hvarrho$. For example, in the QSL based on the relative 
purity~\cite{qmspeedopen} - defined in (\ref{relpurity}) - and for Fubini--Study distance of pure 
states $ds^2$ in $\Xi$ and $ds'^2$ in $\bm[\hm_s]$ have the following 
form~\cite{hourisqgr,hourisqgrym}: 
\be
ds^2 \equiv \Uplambda ds'^2 = \kappa^2 (\tr (\hvarrho d\hvarrho))^2  \label{dsave} 
\ee

The density matrix $\hvarrho$ should include the contribution of all subsystems, except that of 
the clock, because its contributions is implicit through $d\hvarrho$, which is related to the 
chosen clock and time parameter. Moreover, as the quantum speed limit relations are obtained 
from uncertainty relations for continuous parameters, we trace out the contribution of 
discrete states of internal symmetry $G$ of subsystems. Therefore, in general $\hvarrho$ is mixed. 
An extension of the effective path of the state in its Hilbert space including the contribution of 
$G$ symmetry is discussed in~\cite{hourisqgr}. Note also that operators $\hvarrho$, $d\hvarrho$, 
the distance function $\dm (\hvarrho, \hvarrho + d\hvarrho)$, and the metric ${\tt g}_{tt}$ are 
dimensionless. By contrast, comparison of diffeo-surface areas introduces a dimensionful parameter 
in $\Xi$. The constant scale $\Uplambda$, which can be chosen to be proportional to $M_P^{-1}$ (or 
equivalently $L_P$), transforms the dimensionless path parameter $s'$ in the Hilbert space to the 
path parameter $s$ with length dimension in $\Xi$. 

Due to the similarity of $ds$ to affine separation in Riemannian geometry we define an average or 
effective path\footnote{The expression in the r.h.s. of (\ref{dsave}) depends on the QSL inequality 
used for obtaining (\ref{dsave}), which in turn depends on the purity of state $\hvarrho$ and type 
of its evolution. If it consists of a trace operation including the density operator, it can be 
interpreted as averaging and the path as an average. Otherwise, the path would be a representative 
and should be called effective path.} for the subsystems in their parameter $\Xi$, 
see~\cite{houriqmsymmgr,hourisqgrhighlight} for more details. Specifically, it is possible to 
define a one-to-one map between the path of state $\hvarrho$ in the Hilbert space of subsystems 
$\hm_s$ - more precisely $\bm[\hm_s] \owns \hvarrho$ and an effective/average path in $\Xi$, 
up to a constant scaling. As the Hilbert space is projective, the scaling is arbitrary. Thus, we 
can use the same affine separation for both paths, and the effective path in $\Xi$ can be written 
as:
\be
ds^2 = g_{\mu\nu} (X) dX^\mu dX^\nu, \mu,~\nu = 0, \cdots, 3  \label{effmetric}
\ee
where $X^\mu$ are average or effective value of parameters. As the variation of state in the r.h.s. 
of (\ref{dsave}) is defined with respect to a chosen clock, we use the same clock in the 
definition of effective coordinates $X^\mu$. Of course, the time coordinate $X^0$ can be 
reparameterized or redefined arbitrarily. However, for the sake of consistency, one has to follow 
its relation with the chosen clock and reference subsystems used for the definition of relative 
variation of the state $d\hvarrho$. This is analogous to the necessity of choosing a reference 
frame for the definition of classical spacetime coordinates.

In~\cite{houriqmsymmgr} it is demonstrated that signature of the average metric $g_{\mu\nu} (X)$ 
is negative and the effective geometry defined by (\ref{effmetric}) is Lorentzian. Therefore, 
in contrast to $\upeta_{\mu\nu}$ which is arbitrary, $g_{\mu\nu} (X)$ can be identified with the 
metric of the perceived classical spacetime. This observation justifies the claim that the 
perceived classical spacetime and its geometry are emergent concepts. 

Equations (\ref{dsave}) and (\ref{effmetric}) show that $g_{\mu\nu} (X)$ is indeed related to the 
quantum state of subsystem. However, similar to general relativity, (\ref{dsave}) and 
(\ref{effmetric}) do not determine $g_{\mu\nu} (X)$. Indeed, the only truly measurable is $ds$ 
(or equivalently $ds'$), because in \sqgr~there is no spacetime in the classical sense. The metric 
$g_{\mu\nu} (X)$ is only an association by definition and has a physical interpretation as geometry 
only in the classical limit, where it can be calculated by applying variational principle to the 
classical limit of the Lagrangian (\ref{yminvarsub}), namely (\ref{classicgr}) discussed in the 
next section. The Einstein equation obtained in this way corresponds to the constraint 
(\ref{lagrangiangcond}) and ensures consistency of \sqgr.

\subsubsection{Classical Limit} \label{app:classiclimit}
As we remarked in Sec. \ref{app:qm} the standard procedure for finding classical limit of a 
quantum model, namely considering $\hbar \rightarrow 0$ completely destroys the structure 
of \sqgr. On the other hand, when the resolution of experiments is not sufficient for detecting 
quantum effects, we usually neglect quantum nature of objects and physical process and treat 
them classically. This is exactly the situation regarding quantum gravity in current experiments. 
Therefore, regarding pure gravity term, that is pure $\suinf$ part in the action $\lm_{U_s}$ - the 
first term in the integrand of (\ref{yminvarsub}) - its quantum structure is not discernible by 
present experiments, and to experimenter it appears simply as a scalar term depending on (3+1) 
parameters, interpreted as coordinates of a classical {\it spacetime'}. According to a theorem 
in the Riemannian geometry~\cite{curvaturfunc} any scalar function on a Riemannian manifold 
of $d > 2$ is the scalar curvature for some metric associated ti the manifold. For this reason, in 
what concerns gravity, in classical limit the Lagrangian (\ref{yminvarsub}) is perceived as:
\be
{\mathcal L}_{U_s} ~ \autorightarrow{Classical}{limit} ~ {\mathcal L}_{cl.gr} = 
\int_{\Sigma} d^4X \sqrt{|g|} \biggl [\kappa R^{(4)} + \frac{1}{4} \tr (G^{\mu\nu} G_{\mu\nu}) + 
\frac {1}{2} \sum_s \tr (\sD \hrho_s) \biggr ] \label{classicgr}
\ee

The Lagrangian ${\mathcal L}_{cl.gr}$ has the form of the Yang-Mills QFT for local $G$ symmetry 
in a curved space in which the effective metric $g_{\mu\nu}$ is obtained from solving Einstein 
equation obtained from the effective scalar curvature term in (\ref{classicgr}). Moreover, the 
constraint equation (\ref{lagrangiangcond}) becomes equal to the Einstein equation. Considering 
equation (\ref{enermomequilib}) that describes the metric constraint as a null total energy, it is 
clear that the geometric part of the Einstein equation should be considered as gravitational 
energy. These results confirm that gravitational waves are indeed time varying and propagating 
part of the gravitational energy, and in contrast to general relativity we do not have to 
verify gauge invariance~\cite{gwgaugeinvar} and propagation~\cite{gwenertrans} of gravitational 
waves, and other technical issues, see e.g.~\cite{gwhistory} for a historical review of these 
topics. 

\section{A Brief Review of Stable Non-Thermal States in Quantum Many-Body Systems}  \label{app:mbnonthermal}
Isolated many-body quantum systems in a non-equilibrium state have the tendency to thermalize. 
This well known behaviour is formulated as a conjecture called {\it Eigenstate Thermalization 
Hypothesis (ETH)}~\cite{eth}(and references therein). Although ETH is confirmed in many 
condensed matter systems, deviations from it have also been observed\cite{ethviol}(review). 
The first and the best understood example of such phenomena is Many-Body Localization 
(MBL)~\cite{mblrev,mblrev0} (review), in which at thermodynamic limit, that is at 
$t \rightarrow \infty$, finite subregions of the system do not thermalize, and keep a memory 
of their initial state and properties. Another example of the ETH violation occurs when the 
density matrix of a system is divided to non-thermal Krylov subspaces, which evolve 
(approximately) independent of other subspaces - clusters. Moreover, their number increases 
rapidly with the size of the system. In particular, in some systems and for specific initial 
conditions, the number of Krylov subsystems increases exponentially with the system's size. 
Such behaviour is called (strong) {\it Hilbert Space Fragmentation 
(HSF)}~\cite{qmfragrev,qmfragrev0}. Here we extend HSF definition to cases where the state of 
a system evolves to many (approximately) independent independent Krylov subspaces. The 
reason is that we are only interested in the quantum state of the Universe, which according 
to the postulates of \sqgr~its Hilbert space is infinite dimensional.

The origin of HSF which often occurs in constrained and random unitary 
systems~\cite{qmfragstat,qmfragalgebra} is not fully understood. In contrast to MBL it does not 
necessarily correspond to subregions of the lattice (physical space )~\cite{qmfragstat} or 
superselection and factorization of subspaces with conserved charge associated to 
symmetries~\cite{qmfragalgebra}. The Krylov subsystems of interest are generated by unitary 
evolution, that is the application of unitary operator $\hU = \exp(-it\hH)$, where  the initial 
state $\hrho_0$ evolves according to the Hamiltonian $\hH$. The (strong) HSF occurs when 
the number of Krylov subspaces grows exponentially with the size of the system. According 
to~\cite{qmfragalgebra} commutant algebra of the system, that is the algebra of operators 
commuting with all terms in the Hamiltonian can be used to investigate whether HSF can 
occur in a given system. Specifically, the number of Krylov subspaces is equal to the dimension 
of maximal Abelian subalgebra of the commutant algebra. Nonetheless, the actual occurrence of 
HSF depends also on the initial state~\cite{qmfragstat}.

Another concept that may provide insight into Hilbert space fragmentation and evolution of 
the state to an approximate product state is Anderson Orthogonality Catastrophe 
(AOC)~\cite{orthcatast}. Specifically, Anderson showed that the overlap between initial 
$|\psi_0\rangle$ and quenched $|\psi_q\rangle$ ground states of a many-body quantum 
system with $N$ elementary components $||\langle\psi_0|\psi_q\rangle|| \propto 1/N$. Thus, 
irrespective of the strength of the quench and details of the system, for $N \rightarrow \infty$ 
these states are orthogonal to each others. More recently it is 
demonstrated~\cite{orthocatastqmspeed} that AOC is a manifestation of the Quantum Speed 
Limit (QSL) briefly reviewed in Sec. \ref{app:effmetric} and is used to relate the the perceived 
classical spacetime to quantum state of the contents of the Universe. Asymptotic orthogonality 
of states means that their overlap goes to zero.

\subsection {Fragmentation of Hilbert Space of the Universe}  \label{sec:sqgrfrag}
Although the Universe is the only truly isolated quantum system, methods used for studying 
HSF and AOC cannot be directly used in \sqgr~to demonstrate the division of its state to 
product states of approximately isolated subsystems. The reason is the fact that HSF, MBL, 
and AOC are phenomena observed in composite systems consisting of qubits or qudits. 
Moreover, as discussed in sections \ref{app:evoluniv} and \ref{sec:enermominterprt}, in 
\sqgr~the whole Universe is static and there is no Hamiltonian to generate a time evolution 
and its associated Krylov subspaces. Thus, all states are equivalent up to a $\suinf$ 
transformation. Nonetheless, as we briefly discussed in Sec. \ref{app:subsys}, local differences 
between states, that is in subspaces of the Hilbert space $\hm_U$ are meaningful and 
measurable with respect to each others. According to the definition of 
subsystems~\cite{sysdiv} these local features - clusters inside the global state - can be 
treated as approximately isolated subsystems. In fact, even in condensed matter, emergence 
of the HSF is accompanied with the emergence of non-local, but statistically non-trivial 
clusters~\cite{qmfragstat}. This means that in analogy with \sqgr, the global aspects of 
many-body states are involved in the emergence of HSF.

In~\cite{hourisqgr} we showed coherence reduction of the completely coherent state under 
application of a generator of $\suinf$. In Sec. \ref{app:decohere} we further quantified this 
process, which can be interpreted as emergence of features in the global state. As generators 
$\hL^a (\eta, \zeta)$ of $\suinf$ constitute a basis for $\bm[\hm_U]$, any operator 
$\hO \in \bm[\hm_U]$ can be expanded with respect to them. Thus, the results obtained in 
Sec. \ref{app:decohere} can be easily extended to a general operator $\hO$.

Another quantity that heralds and quantifies non-local clustering of a state is the overlap 
between an initial state and its modification after being affected by one or multiple 
operators $\hO \in \bm[\hm_U]$. Note that in the absence of subsystems at this stage, and thereby 
the absence of a quantum clock and time parameter, our approach has to be operational, that is 
we only consider effect of algebraic operations without associating them to a Hamiltonian. In 
Sec. \ref{app:stateoverlap} we calculate the overlap between a state and its modified form 
after multiple applications of $\suinf$ generators. The result shows that fluctuations of the 
state at small scales in continuous parameters $(\eta, \zeta)$ would be amplified. Specifically, 
because generators $\hL^a (x), x \equiv (\eta, \zeta) \in D_2$ depend on the derivative with 
respect to $x$, after $n$ application of $\hL^a$'s, the overlap $\chi$ will depend on $n$ 
derivatives of the wavefunction $\gm (x)$, see Sec. \ref{app:stateoverlap} for definitions and 
details. Consequently, small-scale variations of $\gm (x)$ will be amplified. On the other hand, 
operations restricted to small scales leave the large scale features unchanged. Hence, small 
scale features can be considered as approximately isolated subsystems. In this case, local 
symmetries are those which affect only subspaces associated to these subsystems.

\section{Quantification of Decoherence of Completely Coherent States}  \label{app:decohere}
Various measures for quantification of the coherence of quantum states are proposed in the 
literature~\cite{coheremeasure}. Here we use one of the most generally applicable measures, 
called {\it $\ell_p$-norm} with $p=1$, namely {\it $\ell_1$-norm}. Its monotonously is 
demonstrated in~\cite{coheremeasure}. It quantifies the coherence of a decomposed density 
matrix $\hrho = \sum_{i,j} \rho_{ij} |i\rangle\langle j|$, where $\{|i\rangle\}$ is an orthogonal 
basis, as:
\be
C_{\ell_1} (\hrho) \equiv \sum_{i \neq j} |\rho_{ij}| \label{coherelp1}
\ee
In \sqgr~the state of the whole Universe $|\psi\rangle \in \hm_U$ is characterized by two 
continuous parameters $(\theta, \phi)$ and can be expanded as:
\bea
|\psi \rangle & = & \nm \int d^2\Omega f (\theta, \phi) |\theta, \phi \rangle, \quad \quad 
\nm \int d^2\Omega |f (\theta, \phi)|^2 =1. \label{ustate} \\
\hrho & = & |\psi \rangle \langle \psi | = |\nm|^2 \int d^2\Omega d^2\Omega' f (\theta, \phi) 
f^* (\theta', \phi') |\theta, \phi \rangle \langle \theta', \phi'|  \label{udensity} \\
\tr \hrho & = & |\nm|^2 \int d^2\Omega d^2\Omega' f (\theta, \phi) f^* (\theta', \phi') 
\delta^2 (\Omega - \Omega') \nonumber \\
& = & |\nm|^2\int d^2\Omega |f (\theta, \phi)|^2 =1
\eea
where $\nm$ is a normalization constant and:
\be
\delta^2 (\Omega - \Omega') \equiv \frac{1}{\sqrt{|g^{(2)}|}} \delta (\theta -\theta') 
\delta (\phi -\phi')  \label{deltaomega}
\ee

The $\ell_1-norm$ coherence of $\hrho$ is measured as:
\be
C_{\ell_1} (\hrho) = |\nm|^2 \biggl (\int d^2\Omega d^2\Omega' |f (\theta, \phi) 
f^* (\theta', \phi')| - \int d^2\Omega |f (\theta, \phi)|^2 \biggr )  \label{sqgrcohere}
\ee

In a completely incoherent state $\hvarrho_{ci}$ all off-diagonal components are 
null. Thus, as expected, $C_{\ell_1} (\hvarrho_{ci}) = 0$. By contrast, a completely 
coherent state corresponds to $f (\theta, \phi) = 1$ and 
$\nm = (\int d^2\Omega)^{-1} = 1/4\pi$. Thus:
\be
\hvarrho_{cc} = \frac{1}{(4\pi)^2} \int d^2\Omega d^2\Omega' |\theta, \phi \rangle 
\langle \theta', \phi'|  \label{ccdensity}
\ee

Using (\ref{sqgrcohere}) and the fact that diagonal terms which are eliminated by the 
second term constitute a zero measure, we find $C_{\ell_1} (\hrho_{cc}) =1$. All other states 
have a smaller coherence measure $C_{\ell_1}$~\cite{coheremeasure}.Application of an 
operator $\hO \in \bm[\hm_U]$ to $\hvarrho_{cc}$ changes it to 
$\hrho_{cc} \rightarrow \hO \hrho_{cc} \hO^\dagger \neq \hrho_{cc}$. Due to the monotonously 
of norm-${\ell_1}$ the new state necessarily has $C_{\ell_1} < 1$. Thus, it is less coherent.

\section{Evolution of Overlap Between States}  \label{app:stateoverlap}
One way to quantify fragmentation of the Hilbert space and emergence of non-thermal features, 
reviewed in Sec. \ref{app:mbnonthermal}, is comparing an initial state $|\psi_0\rangle$ with 
its evolved form under application of an operator $\hO \in \bm[\hm_U]$, that is 
$|\psi_1\rangle = \hO |\psi_0\rangle$. The quantification is through the calculation of their 
projection or in other word overlap $\chi$ defined as 
$\chi \equiv ||\langle \psi_0 |\psi_1\rangle|| = (\tr(\varrho_0 \varrho_1))^{1/2}$. The overlap 
appears in several quantum information measures, for instance in the definition of relative 
purity (\ref{relpurity}) and in the Bures angle $\Theta_B \equiv \arccos \chi$ and related QSL 
for mixed states, see e.g.~\cite{qmspeedrev} for a review. In particular, $\chi \rightarrow 0$ 
corresponds to the evolution of initial state to a completely distinct state.

Generators of $\suinf$ constitute a basis for $\bm[\hm_U]$. Therefore, operators 
$\hO \in \bm[\hm_U]$ can be expanded with respect to $\hL_a(x),~ x \equiv (\eta, \zeta) \in D_2$: 
$\hO = \exp(i \sumint_a d\Omega^{(2)}\epsilon_a (x) \hL_a (x))$, where $d\Omega^{(2)}$ is defined 
in (\ref{lagrange2d}). Because the action $\lm_U$ is topological and we are interested in local 
modification of states, we use directly the above expansion of operators rather than Feynman 
diagrams generated by the action. However, despite apparent simplicity of the expansion of $\hO$, 
non-commutativity of $\hL_a$ makes the calculation of overlap lengthy. For this reason, we only 
consider the application of two generators on an arbitrary state and use the results to deduce 
the effect of a larger number of applied generators. 

The overlap under application of one and two operators can be written as:
\bea
\chi_1^2 & = & \tr(\hvarrho_0 \hO_1 \hvarrho_0 \hO_1^\dagger)  \label{overlapop1} \\
\chi_2^2 & = & \tr(\hvarrho_0 \hO_2 \hO_1 \hvarrho_0 \hO_1^\dagger \hO_2^\dagger)  \label{overlapop2}
\eea

In \sqgr~vectors of the Hilbert space $\hm_U$ are complex valued functions 
$|\psi_0 \rangle = \gm (x), ~x \in D_2$, where $D_2$ is diffeo-surface of $\suinf$ symmetry 
represented by $\hm_U$. We are interested in the variation of the arbitrary state $\gm (x)$ 
under successive application of $\suinf$ generators $\hL_a$. For this purpose it is more 
convenient to use the torus basis~\cite{suinftriang,suinftriang0,suninftorus} 
(see ~\cite{hourisqgrym} for a brief review). We remind that all foliations and orthogonal 
functions used for the definition of $\suinfa$ algebra are locally equivalent and can be used 
irrespective of the topology of diffeo-surface. In torus basis generators, their multiplication, 
and their Lie algebras are formulated as the followings:
\vspace{-9pt}
\bea
&& \hL_{\boldsymbol{m}} = -i e^{i\boldsymbol{m}.x} (\boldsymbol{m} \times \partial_x), \quad \quad 
\boldsymbol{m} = (m_1, m_2), ~ m_1, m_2 \in \mathbb{Z}, \quad \quad 
\boldsymbol{m} \times \boldsymbol{n} = m_1n_2 - m_2n_1  \label{suinfgentorus}  \\
&& \hL_{\boldsymbol{m}} \hL_{\boldsymbol{n}} = - e^{i(\boldsymbol{m+n}).x} 
\biggl (i (\boldsymbol{m \times n}) + \boldsymbol{m} \times \partial_x \biggr ) 
(\boldsymbol{n} \times \partial_x)  \label{torusgenmultip} \\
&& [\hL_{\boldsymbol{m}}, \hL_{\boldsymbol{n}}] = (\boldsymbol{m} \times \boldsymbol{n}) 
\hL_{\boldsymbol{m+n}}  \label{torusalgebra}
\eea

Using (\ref{overlapop1}) with $\hO_1 = \hL_{\boldsymbol{m}}$, we find: 
\bea
\chi_1^2 (\boldsymbol{m}) & = &\biggl (\int d^2\Omega \gm^* e^{i\boldsymbol{m}.x} \boldsymbol{m} 
\times \partial \gm \biggr ) \biggl (\int d^2\Omega \gm^* e^{-i\boldsymbol{m}.x} \boldsymbol{m} 
\times \partial \gm \biggr )  \nonumber \\
& = & \biggl (\boldsymbol{m} \times \fm_{\boldsymbol{m}} [\gm^* \partial \gm] 
\biggr) \boldsymbol{.} \biggl (\boldsymbol{m} \times \fm_{(\boldsymbol{-m})} 
[\gm^* \partial \gm]\biggr)  \label{overlapfourier1}
\eea
where $\fm_{\boldsymbol{m}}[f]$ means mode $\boldsymbol{m}$ of 2D Fourier transform of function $f$. 
Properties of Fourier transform can be used to make the expression of $\chi_1$ more concise:
\be
\chi_1 (\boldsymbol{m}) = \biggl | \sum_{\boldsymbol{l}} (\boldsymbol{m \times l}) ~ 
G^*_{\boldsymbol{m+l}} ~ G_{\boldsymbol{l}} \biggr | \equiv \sum_{\boldsymbol{l}} \am_{\boldsymbol{l}}  
\label{chi1fourier}
\ee
where $G_{\boldsymbol{l}}$'s are Fourier coefficients of $\gm (x)$. From (\ref{chi1fourier}) we 
can immediately conclude that modes parallel to $\boldsymbol{m}$ do not contribute in 
$\chi_1 (\boldsymbol{m})$. We can also estimate the contribution of modes much larger or 
smaller than $|\boldsymbol{m}|$. Specifically, the wavefunction $\gm (x)$ has to be smooth - 
at least its first and second derivatives must be finite. In addition it must be $L^{(2)}$ 
integrable. Under these conditions the amplitude of its Fourier modes asymptotically decreases 
with increasing mode number after a limit mode $\boldsymbol{l}_0$, see e.g.~\cite{fourieranal}. 
Thus, for $|\boldsymbol{l}| \gtrsim \boldsymbol{l}_0$:
\vspace{-9pt}
\bea
&& \am_{\boldsymbol{l}} \lesssim \frac{C^2 |\boldsymbol{m \times l}|}
{|\boldsymbol{m+l}||\boldsymbol{l}|} \lesssim \frac{C^2 |\boldsymbol{m}|}{|\boldsymbol{m+l}|}, 
\quad \quad |\boldsymbol{l}| \gtrsim \boldsymbol{l}_0  \label{chimodeampapprox} \\
&& 
\begin{cases}
\am_{\boldsymbol{l}} \lesssim C^2 & \text{For} ~ |\boldsymbol{l}| \ll |\boldsymbol{m}| \\
\am_{\boldsymbol{l}} \lesssim \frac{C^2 |\boldsymbol{m}|}{|\boldsymbol{l}|} \rightarrow 0 & 
\text{For} ~ |\boldsymbol{l}| \gg |\boldsymbol{m}| 
\end{cases} \label{amplimit}
\eea
where $C$ is a constant. This simple estimation shows that an operator - here a generator of 
$\suinf$ symmetry - affects some modes and leave others approximately untouched. According to 
(\ref{amplimit}) modes much larger than $|\boldsymbol{m}|$ - in other words UV modes - are more 
affected and contribute less to the overlap of the two states. In other words the two states 
are more different at small scales (in the Hilbert space). This is a direct consequence of the 
presence of a derivative in generators $\hL_{\boldsymbol{m}}$. The emergence of UV clustering means 
that local interactions are stronger than global $\suinf$ (gravity) interaction. Thus, 
\sqgr~is consistent with Weak gravity conjecture~\cite{weakgrconj}.

Using (\ref{torusgenmultip}) we also calculate $\chi_2^2$:
\vspace{-9pt}
\bea
\chi_2^2(\boldsymbol{m}, \boldsymbol{n}) & = & \biggl (\int d^2\Omega e^{i\boldsymbol{m+m}.x} 
(\boldsymbol{n} \times \partial \gm^*) . (\boldsymbol{m} \times \partial \gm) \biggr ) 
\biggl (\int d^2\Omega e^{-i\boldsymbol{m+m}.x} (\boldsymbol{m} \times \partial \gm^*) . 
(\boldsymbol{n} \times \partial \gm) \biggr )  \nonumber \\
&& \biggl | \sum_{\boldsymbol{l}} (\boldsymbol{m} \times \boldsymbol{l}.(\boldsymbol{n} \times 
(\boldsymbol{m+n+l})) G^*_{\boldsymbol{l}} ~ G_{\boldsymbol{m+n+l}} \biggr |^2 \equiv \biggl | 
\sum_{\boldsymbol{l}} B_{\boldsymbol{l}} \biggr |^2  \label{overlapfourier2}
\eea

An analysis of asymptotic behaviour of modes as performed for $\chi_1$ shows that for 
$|\boldsymbol{l}| \gtrsim \boldsymbol{l}_0$, the amplitude of $B_{\boldsymbol{l}}$ modes is roughly 
independent of $|\boldsymbol{l}|$ and is determined solely by $|\boldsymbol{m}|$ and 
$|\boldsymbol{n}|$: $B_{\boldsymbol{l}} \lesssim C'|\boldsymbol{m}||\boldsymbol{n}|$. More generally
(\ref{overlapfourier2}) shows that the overlap is dominated by UV modes, because each new 
generator increases the order of derivatives in the overlap by 1. This is not in contradiction 
with (\ref{chimodeampapprox}) and (\ref{amplimit}). After multiple application of generators, 
small scale features become randomized and differences are smeared, leading to larger 
contribution to the overlap. By contrast, at large scales the overlap decreases and states are 
approximately orthogonal to each other.

\end{document}